\documentstyle[aps,12pt,eqsecnum,epsfig]{revtex}

\begin{document}

\title{
\begin{flushright}
{\normalsize SLAC-PUB-7550}\\
{\normalsize DOE/ER/40561-329-INT97-20-03}\\
{\normalsize DOE/ER/41014-14-N97 }\\
\hspace {3.7in} June 1997\\
\end{flushright}
  Light Front   Treatment of Nuclei: Formalism and Simple Applications}
\author{Gerald A. Miller\footnote{ 
permanent address Department of Physics, Box 351560, University of Washington,
Seattle, WA 98195-1560}  
   \\Stanford Linear Accelerator Center, Stanford University,
Stanford, 
California 94309   
\\national Institute for Nuclear Theory, Box 35150, University of Washington,
Seattle, WA 98195-1550
 \vspace{15pt} }

\maketitle 

\begin{abstract}
A relativistic light front treatment of nuclei is developed by
 performing light front quantization for a chiral Lagrangian.  The
 energy momentum tensor and the appropriate Hamiltonian are obtained.
 Three illustrations of the formalism are made.  (1) Pion-nucleon
 scattering at tree level is shown to reproduce soft pion theorems.
 (2) The one boson exchange treatment of nucleon-nucleon scattering is
 developed and shown (by comparison with previous results of the equal time
 formulation) to lead to a reasonable description of nucleon-nucleon
 phase shifts.  (3) The mean field approximation is applied to
 infinite nuclear matter, and the plus momentum distributions of that
 system are studied. The mesons are found to carry a significant
 fraction of the plus momentum, but are inaccessible to experiments.

\end{abstract}
\pacs{21.30.Fe  21.65.+f  24.10.Jv  13.60.Hb  11.80.-m 13.75.-n 13.75.Cs
03.65.Pm   }
\newpage

\section{Introduction}

The need for a relativistic methodology that is broadly applicable to nuclear
physics has never been more apparent. 
One of our  most important sets of problems involves understanding
the transition between the (baryon, meson)
and the (quark, gluon) degrees of freedom.
Using a relativistic
formulation of the hadronic degrees of freedom is necessary to avoid a 
misinterpretation of a kinematic effect as a signal for the transition.

The goal of understanding future high momentum studies
of nuclear targets using 
 exclusive, nearly exclusive or inclusive processes can only be met through
using 
relativistic techniques. The light front approach of Dirac\cite{Di 49}
 in which the
time variable is taken as $t+z$ and the spatial variables 
are  $t-z,x,y$
\cite{lcrevs,hari}
is one of the promising approaches because
the momentum canonically conjugate to $t-z$, $p^+\equiv p^0+p^3$, is directly
related to the observables. 

It is worthwhile to begin with a qualitative explanation
of  the utility of these light cone variables and
this light front approach in a qualitative fashion.
 Consider lepton-nucleus
 deep inelastic scattering as a first example. The observed 
 structure function 
depends on the Bjorken
variable 
$x_{Bj}$ which in the parton model 
 is the ratio of the quark plus momentum to that of the target. If one regards
the nucleus as a collection of nucleons, $x_{Bj}
=p^+/k^+$, where  $k^+$ is the 
plus momentum of a nucleon bound in the nucleus.
Thus, a more 
direct relationship between the necessary 
nuclear  theory and experiment occurs by
using a theory in which $k^+$ is one of the canonical variables.
Since $k^+$ is conjugate to a spatial variable  $x^-\equiv t-z$, 
it is natural to quantize the dynamical variables 
at the equal light cone time variable of $x^+\equiv t
+z$. To use such a formalism is to use  light front quantization, 
since the other three spatial coordinates ($x^-,\bbox x_\perp$)
are  on a plane perpendicular to a  light like vector\cite{notation}. 
This use of light front quantization 
requires a new derivation of  
the nuclear wave 
function,   because previous work used
  the equal time formalism.     

Are these light cone variables useful only in nuclear deep 
inelastic scattering? Let's answer this by examining the origin of 
such coordinates. The
four momentum of the incident virtual photon, q, can be said to have the
components $q=(\nu,0,0,-\nu- Q^2/2\nu)$, with $q^2=-Q^2$, and $\nu, Q^2$
very large but $Q^2/\nu$ finite (the Bjorken limit).
Then $x_{Bj}\equiv {Q^2\over 2 k\cdot q}={Q^2\over k^+q^-}$ 
The condition that the reaction be elastic scattering from the  quarks
is that $(p+q)^2=p^2, $ or $2p\cdot q=Q^2=p^+q^-$. Thus 
$x_{Bj}
=p^+/k^+$ results from having only one  large momentum in the problem, which
can be taken in the negative z-direction, so that minus component is enhanced.
More generally, one expects to be able to use light cone coordinates 
$(p^+,p^-,p_\perp)$ whenever there is such a large momentum in the problem
as in 
any high energy scattering process.  Diverse applications
are shown in  the
text by Cheng and Wu\cite{chengwu}.
 Examples of most relevance  
include  high energy projectile nuclear scattering and 
high momentum transfer quasi-elastic reactions involving nuclear
targets.

Light front techniques have previously been applied to systems of
two hadrons. The two main approaches have been the relativistic
quantum mechanics of directly interacting particles \cite{coester,Ke
91,Ka 88,Fu 91} and relativistic field theory
\cite{lcrevs,fs1,fs2}. We choose here to employ specific Lagrangians
which embody chiral and other symmetries, and thus use field theory.

The light front quantization procedure necessary to treat nucleon
interactions with scalar and vector mesons was derived by
Soper\cite{des71}, and by Yan and
collaborators\cite{yan12,yan34}. Here we combine the previous
formalisms to obtain a light front treatment of a Lagrangian which
contains pions, vector and scalar mesons, and which respects the
constraints of chiral symmetry.

Here is an outline. The 
bulk of the formalism is presented in Sect. 2. First, a
chiral Lagrangian which includes pions,
scalar mesons and neutral
vector mesons is presented. The field equations are derived, and
the quantization procedure for 
the free and interacting fields are quantized at the zero of light cone
time $x^+$. 
The energy
momentum tensor, the light front hamiltonian $P^-$ and 
plus momentum operator $P^+$ are derived. The necessary contact interactions
involving the exchange of
instantaneous fermions and vector bosons are obtained.
The principal 
purpose of the present work is to develop a technique that could have
wide application in nuclear physics. Thus we 
study and check the present formalism by applying it to  three
different examples- $\pi N$ and $NN$ scattering, and a mean field
treatment of infinite nuclear matter- of relevance to nuclear physics.
Dealing successfully with 
each of these subjects is a prerequisite for making progress.

Sect. 3 shows how light front field theory leads to a chiral treatment
of low energy pion-nucleon scattering, which is consistent with the
results of soft pion theorems. Then nucleon-nucleon scattering is
handled in a manifestly covariant manner, within the one-boson
approximation, in Sect. 4.  A discussion of the impact of chiral
symmetry on the two-nucleon intermediate state contribution 
to the two pion exchange potential is also included.  The
mean field approximation is applied to infinite nuclear matter in Sect
5.  Glazek and Shakin\cite{gs} used a Lagrangian containing nucleons
and scalar mesons to study infinite nuclear matter. Here vector mesons
are included and the rotational-invariance arguments used in Sect. 4 
are used to derive the  Glazek-Shakin $k^+$ variable. 
The energy of nuclear matter is computed and shown to be the
same as found in the equal time formalism. The unique feature of the present 
formalism is the ability to obtain the nuclear and mesonic plus-momentum 
distributions from the energy momentum tensor.  We find that that mesons
can carry a significant fraction of the nuclear plus momentum, but have
support only at 0 plus-momentum.  
Some the results for nuclear matter  have
been presented in an earlier publication\cite{me97}; here the calculation
is performed in two different ways and explained in more detail.  
Sect. 6 summarizes the new results, presents a critique and discusses
possible future applications. Appendix A contains a summary of
notation and some useful equations. 

\section {Light Front Quantization}

\subsection{Lagrangian and Field equations}

We use  a non-linear chiral
model in which the 
nuclear constituents are nucleons $\psi$ (or $\psi')$, pions $\bbox{\pi}$
 scalar mesons $\phi$\cite{scalar} and
 vector mesons
$V^\mu$. 
The  Lagrangian ${\cal L}$ is given by 
\begin{eqnarray}
{\cal L} ={1\over 2} (\partial_\mu \phi \partial^\mu \phi-m_s^2\phi^2) 
-{1\over  4} V^{\mu\nu}V_{\mu\nu} +{m_v^2\over 2}V^\mu V_\mu \nonumber\\
+{1\over  4}f^2Tr (\partial_\mu\;U\;\partial^\mu\;
U^\dagger)+{1\over  4}m_\pi^2f^2\;
Tr(U +U^\dagger-2)
+\bar{\psi}^\prime\left(\gamma^\mu
({i\over 2}\stackrel{\leftrightarrow}{\partial}_\mu
-g_v\;V_\mu) -
MU -g_s\phi\right)\psi' \label{lag}
\end{eqnarray}
where the bare masses of the nucleon, scalar and vector mesons are given by 
$M, m_s,$  $m_v$, and  $V^{\mu\nu}=
\partial ^\mu V^\nu-\partial^\nu V^\mu$. The unitary  matrix $U$ can be
chosen from amongst three forms $U_i$:
\begin{equation}
U_1\equiv e^{i  \gamma_5 \bbox{\tau\cdot\pi}/f},\quad
U_2\equiv{1+i\gamma_5\bbox{\tau}\cdot\bbox{\pi}/2f\over
1-i\gamma_5\bbox{\tau}\cdot\bbox{\pi}/2f},\quad
U_3=\sqrt{1-\pi^2/f^2}+i\gamma_5\bbox{\tau\cdot\pi}/f, \label{us}
\end{equation}
which correspond to different definitions of the fields.

The pion-nucleon  coupling here is chosen as that of  linear representations
of chiral symmetry used by  Gursey \cite{gursey}, with the 
the Lagrangian approximately $(m_\pi\ne 0$)
invariant under the chiral transformation
\begin{eqnarray}
\psi^\prime\to e^{i \gamma_5 \bbox{\tau}\cdot\bbox{a}}\psi^\prime\nonumber\\
U\to e^{-i \gamma_5 \bbox{\tau}\cdot\bbox{ a}} \;U\; 
e^{-i \gamma_5 \bbox{\tau}\cdot \bbox{a}}.
\label{chiral}
\end{eqnarray}
One may transform the fermion fields, by taking $U^{1/2}\psi'$ as the
nucleon field. One then gets Lagrangians of the non-linear
representation, as explained by Weinberg \cite{weinberg}. In this case
the early soft pion theorems are manifest in the Lagrangian, and the
linear pion-fermion coupling is of the pseudovector type.  However,
the use of light front theory, requires that one find an easy way to
solve the constraint equation that governs the fermion field. We shall
show that the constraint can be handled in a simple fashion by using
the linear representation. Moreover, we shall see that the early soft
pion theorems are indeed manifest from the form of the light front
Hamiltonian.

The constant $M\over f$ plays the role of the bare 
pion-nucleon coupling constant.
If $f$ is chosen to be the pion decay constant, the Goldberger-Trieman
relation yields the result that the axial vector coupling constant
$g_A=1$, which would be  a problem for the Lagrangian, unless loop effects
can make up the needed 25\% effect. Corrections of that size 
are typical of order $({M\over f})^3$ effects found in the 
cloudy bag model\cite{cbm} for many observables, including $g_A$.

There are no explicit $\Delta's$ in the above Lagrangian. Those will
be handled in a future publication. For the moment we note that
treating the higher order effects of the pion-nucleon inherent in this
Lagrangian is likely to lead to a resonance in the (3,3) channel of
pion nucleon scattering.  Such effects can be included in the two-pion
exchange contribution to nucleon-nucleon scattering. However, such an
approach seems cumbersome.

The choice of using an explicit $\Delta$ instead of the iterated
$\pi-N$ interaction is analogous to our use of a scalar meson even
though the effects of $\pi-\pi$ interactions, which could lead to
similar effects, are included in the Lagrangian. We follow many
authors (see the review \cite{bsjdw}) and include a scalar meson to
simplify calculations.  In this treatment, which follows that of
Refs.\cite{fu96a,sw97}, the scalar meson $\phi$ is not a chiral
partner of the pion- the chiral transformation is that of
Eq.~(\ref{chiral}).

The present Lagrangian
may be thought of
as a low energy effective
theory for nuclei under normal conditions. 
A more sophisticated Lagrangian is reviewed in \cite{sw97} and used in 
\cite {fu96a}; the present
one is used to show that light front techniques can be applied to
 hadronic theories relevant for nuclear physics.
This hadronic model, when evaluated in mean field approximation, 
gives\cite{bsjdw}  at least a qualitatively 
good description of many (but not all) 
nuclear properties and reactions. There are a variety of problems occuring
when higher order terms are included\cite{sw97}. The aim here is
to use a reasonably sophisticated 
Lagrangian to study the effects that one might obtain by using
a light front formulation.

We could also have used the linear 
sigma model. The light front quantization for that model can
be accomplished using a simple generalization of the work of 
Refs.~\cite{des71} and \cite{yan12},
and is not shown here. According to the review \cite{sw97} the use of
such a Lagrangean precludes a successful description of nuclei at the 
mean-field level.

The next step is to examine the field equations. The relevant Dirac equation
 for the nucleons is 
\begin{eqnarray}
\gamma\cdot(i\partial-g_v V)\psi'=(M\;U+g_s\phi)\psi'. \nonumber\\
\label{dirac}
\end{eqnarray}
The field equations for the mesons are 
\begin{eqnarray}
\partial_\mu V^{\mu\nu}+m_v^2 V^\nu&=&g_v\bar \psi'\gamma^\nu\psi'
\label{vmeson}
\end{eqnarray}
\begin{eqnarray}
\partial_\mu\partial^\mu \phi+m_s^2\phi&=&-g_s\bar\psi'\psi'. \label{smeson}
\end{eqnarray}
\begin{eqnarray}
 \partial_\mu\partial^\mu \pi_i +m_\pi^2f\mbox {sin}(\pi/f){\pi_i\over \pi }
+\partial_\mu\left[{\pi_i\over \pi}\partial^\mu\pi(1-{f^2\over \pi^2}
\mbox{sin}^2{\pi\over f})\right]
=-m\bar\psi'
{\partial U\over \partial \pi_i}\psi', \label{pimeson}
\end{eqnarray}
where $\pi=(\sum_j \pi_j^2)^{1/2}$.

The next step is obtain the 
 light front Hamiltonian ($P^-$) as a sum of a free, non-interacting and
a set of terms containing all of the interactions. This is accomplished by
using the energy momentum tensor as 
\begin{equation}
P^\mu={1\over2}\int dx^-d^2x_\perp\;T^{+\mu}(x^+=0,x^-,\bbox{x}_\perp).
\end {equation}
The usual relations determine  $T^{+\mu}$, with
\begin{equation}
T^{\mu\nu}=-g^{\mu\nu}{\cal L} +\sum_r{\partial{\cal L}\over\partial
  (\partial_\mu\phi_r)}\partial^\nu\phi_r,\label{tmunu}
\end {equation}
in which the degrees of freedom are labelled by $\phi_r$.

\subsection{Free Meson Fields} 
It is worthwhile to consider the limit in which the interactions between the 
fields are removed.  This will allow us to define the free Hamiltonian 
$P^-_0$ and to display  the necessary commutation relations.
The  energy momentum tensors of the non-interacting fields are defined as
as $T_0^{\mu\nu}(\phi),T_0^{\mu\nu}(V),$ and  $T_0^{\mu\nu}(\pi).$
The fermion fields are quantized in the next sub-section.
Then the use of Eq.(\ref{tmunu})  leads to the result
\begin{equation}
T^{\mu\nu}_0(\phi)=\partial^{\mu} \phi\partial^\nu \phi-{g^{\mu\nu}\over2}
\left[\partial_\sigma \phi \partial^\sigma \phi-m_s^2\phi^2\right],
\end {equation}
with
\begin{equation}
T^{+-}(\phi)={1\over 2}\bbox{\nabla}_\perp \phi \cdot \bbox{\nabla}_\perp\phi+
{1\over 2}m^2_s \phi^2.
\end {equation}
The scalar field can be expressed in terms of creation and destruction 
operators:
\begin{equation}
\phi(x)=
\int{ d^2k_\perp dk^+ \theta(k^+)\over (2\pi)^{3/2}\sqrt{2k^+}}\left[
a(\bbox{k})e^{-ik\cdot x}
+a^\dagger(\bbox{k})e^{ik\cdot x}\right],
\end {equation} 
where 
$k\cdot x={1\over2}(k^-x^++k^+x^-)-\bbox{k_\perp\cdot x}_\perp$ with
 $k^-={k_\perp^2+m_s^2\over k^+}$, and $\bbox{k}\equiv(k^+,\bbox{k}_\perp)$.
The $\theta$ function restricts $k^+$ to positive values.
The commutation relations  are
\begin{equation}
[a(\bbox{k}),a^\dagger(\bbox{k}')]=
\delta(\bbox{k}_\perp-\bbox{k}'_\perp)
\delta(k^+-k'^+)\label{comm}
\end {equation}
with $[a(\bbox{k}),a(\bbox{k}')]=0$.
It is useful to  define 
\begin{equation}
\delta^{(2,+)}(\bbox{k}-\bbox{k}')\equiv 
\delta(\bbox{k}_\perp-\bbox{k}'_\perp)\delta(k^+-k'^+),\label{def3d}
\end {equation}
which will be used throughout this paper.

The derivatives appearing in the quantity $T^{+-}$ are evaluated and then one
sets $x^+$ to 0 to obtain the result
\begin{equation}
 P_0^-(\phi)=\int d^2k_\perp dk^+\theta(k^+)a^\dagger
(\bbox{k})a(\bbox{k}){k_\perp^2+m_s^2\over k^+},
\end{equation}
which has the interpretation of an operator the counts the light front
energy $k^-={k_\perp^2+m_s^2\over k^+}$ of all of the particles.

The pion field is treated in a similar manner, with the result
\begin{equation}
\bbox{\pi}(x)=
\int{ d^2k_\perp dk^+ \theta(k^+)\over (2\pi)^{3/2}\sqrt{2k^+}}\left[
\bbox{a}(\bbox{k})e^{-ik\cdot x}
+\bbox{a}^\dagger(\bbox{k})e^{ik\cdot x}\right],
\end {equation} 
and
\begin{equation}
 P_0^-(\pi)=\int d^2k_\perp dk^+\theta(k^+)\bbox{a}^\dagger(\bbox{k})\cdot
\bbox{a}(\bbox{k}){k_\perp^2+m_\pi^2\over k^+},
\end{equation}
with commutation relations analogous to that of Eq.(\ref{comm}).

The energy momentum tensor for the free vector meson field is obtained 
directly from the defining relation (\ref{tmunu}) as
\begin{equation}
T_0^{\mu\nu}(V)=V^{\alpha\mu}\partial^\nu V_\alpha+g^{\mu\nu}\left[{1\over 4}
V^{\alpha\beta}V_{\alpha\beta}-{m_v^2\over 2}V_\alpha V^\alpha\right].
\end{equation}
It is desirable to obtain the symmetric energy momentum tensor. This is done
by using $\partial^\nu V^\alpha=\partial^\alpha V^\nu+V^{\nu\alpha}$,
 subtracting a total divergence and using the free field equations. The result
is
\begin{equation}
T_0^{\mu\nu}(V)=V^{\alpha\mu} V^{\nu\beta}g_{\beta\alpha}+
m_v^2V^\mu V^\nu+g^{\mu\nu}\left[{1\over 4}
V^{\alpha\beta}V_{\alpha\beta}-{m_v^2\over 2}V_\alpha V^\alpha
\right].
\end{equation}
The component relevant for the light front Hamiltonian can be shown 
to be
\begin{equation}
T_0^{+-}(V)={1\over 2}V^{\alpha+}\partial^-V_\alpha-V^{\alpha+}\partial_\alpha
 V_\alpha+{m_v^2\over 2}V^kV^k. \label{t+-}
\end{equation}
The expression for the vector meson field operator is
\begin{equation}
V^\mu(x)=
\int{ d^2k_\perp dk^+ \theta(k^+)\over (2\pi)^{3/2}\sqrt{2k^+}}
\sum_{\omega=1,3}\epsilon^\mu(\bbox{k},\omega)\left[
a(\bbox{k},\omega)e^{-ik\cdot x}
+a^\dagger(\bbox{k},\omega)e^{ik\cdot x}\right],\label{vfield}
\end {equation} 
where the polarization vectors are the usual ones:
\begin{eqnarray}
k^\mu\epsilon_\mu(\bbox{k},\omega)=0,\quad \epsilon_\mu(\bbox{k},\omega)
\epsilon^\mu(\bbox{k},\omega')=-\delta_{\omega\omega'}\nonumber\\
\sum_{\omega=1,3}\epsilon^\mu(\bbox{k},\omega)
\epsilon^\nu(\bbox{k},\omega)=-(g^{\mu\nu}-{k^\mu k^\nu\over m_v^2}).
\end{eqnarray}
Once again the four momenta are on-shell with 
$k^-={\bbox{k}_\perp^2+m_v^2\over k^+}.$ 
The light front commutation relations: 
\begin{equation}
[a(\bbox{k},\omega),a^\dagger(\bbox{k}',\omega')]=\delta_{\omega\omega'}
\delta^{(2,+)}(\bbox{k}-\bbox{k}'), 
\end {equation}
with the others vanishing, lead to commutation relations amongst the
field operators that are the 
same as in Ref.~\cite{yan34}.
The expression for $P_0^-(V)$ can now be obtained from Eqs.(\ref{t+-}) and
(\ref{vfield}) as
\begin{equation}
P_0^-(V)=\sum_{\omega=1,3}\int{ d^2k_\perp dk^+ 
\theta(k^+){\bbox{k}_\perp^2+m_v^2\over k^+}}
a^\dagger(\bbox{k},\omega) a(\bbox{k},\omega).
\end {equation}

\subsection{Interacting Fields}

This subject is complicated by the presence of massive
vector meson fields. Various difficulties
were handled by Soper\cite{des71} 
and Yan\cite{yan34}. The key features that we use  
are summarized here. In particular, 
the fields $V^+,V^{+i}$ are chosen as the three 
independent fields, with the others expressible in terms of these.
We shall need only one of these relationships in which 
the plus component of Eq.(\ref{vmeson})
can be used to obtain 
\begin{equation}
V^{-+}={2\over\partial^+}\left[g_v\;J^+-m^2_vV^+-\partial_iV^{i+}\right].\label{vpm}
\end {equation}
with 
\begin{equation}
J^\mu\equiv \bar\psi'\gamma^\mu\psi'.
\end {equation}
and  the inverse of $\partial^+$ is 
defined in Refs.~\cite{des71}-\cite{yan34}. A more
recent discussion is given by Harindrinath and Zhang\cite{harizhang}, 
and the essentials are presented here in  Appendix A,


We turn to the case of spin 1/2 fermions. Although described by four-component
spinors, these fields have only two independent degrees of freedom. 
The light front formalism allows a convenient separation of
dependent and independent variables via the 
projection operators $\Lambda_\pm\equiv \gamma^0\gamma^\pm/2$\cite{des71}, with
$\psi'_\pm\equiv\Lambda_\pm \psi'_\pm$. The independent Fermion degree of freedom
is chosen to be $\psi'_+$. The properties of the projection operators are
discussed in Appendix A.  One gets two coupled equations for $\psi'_\pm$
by multiplying Eq.(\ref{dirac}) by $\Lambda_+$ and $\Lambda_-$:
\begin{eqnarray}
(i\partial^--g_vV^-)\psi'_+=(\bbox{\alpha}_\perp\cdot
(\bbox{p}_\perp-g_v\bbox{V}_\perp)+\beta (M U+g_s\phi))\psi'_-\nonumber\\
(i\partial^+-g_vV^+)\psi'_-=(\bbox{\alpha}_\perp\cdot
(\bbox{p}_\perp-g_v\bbox{V}_\perp)+\beta (M U+g_s\phi))\psi'_+.
\end{eqnarray}
 The relation between $\psi'_-$ and $\psi'_+$ 
is very complicated unless one may set the plus component of the vector field
to zero\cite{lcrevs}. This is a matter of a choice of gauge for 
QED and QCD, but the non-zero mass of the vector meson prevents such a choice 
here. Instead,
one     simplifies   the equation
for $\psi'_-$ by\cite{des71,yan34}
   transforming  the Fermion field according to 
\begin{equation}\psi'=e^{-ig_v\Lambda(x)}\psi \label{trans}
\end{equation} with 
\begin{equation}
\partial^+ \Lambda=V^+
.\end{equation}
 This 
transformation  leads to the result 
\begin{eqnarray}
(i\partial^--g_v \bar V^-)\psi_+=(\bbox{\alpha}_\perp\cdot 
(\bbox{p}_\perp-g_v\bbox{\bar V}_\perp)+\beta(MU+g_s\phi))\psi_-\nonumber\\
i\partial^+\psi_-=(\bbox{\alpha}_\perp\cdot 
(\bbox{p}_\perp-g_v\bbox{\bar V}_\perp)+\beta(MU+g_s\phi))\psi_+\, \label{yan}
\end{eqnarray}
where
\begin {equation}
 \partial^+\bar V^\mu=\partial^+V^\mu-\partial^\mu V^+= V^{+\mu}. \label{vbar}
\end{equation}
 
 Note that all of the previously obtained Fermionic sources 
of meson fields are unchanged by the transformation (\ref{trans}):
\begin{eqnarray}
\bar \psi\psi=\bar \psi'\psi'\nonumber\\
\bar\psi U\psi=\bar \psi'U\psi'\nonumber\\
\bar\psi\gamma^\mu\psi=\bar\psi'\gamma^\mu\psi'.
\end{eqnarray}

The eigenmode expansion  for $\bar V^\mu$ is needed  to compute the
interaction between 
nucleons. Eqs.~(\ref{vfield})   and (\ref{vbar}) can be used to
obtain
\begin{equation}
\bar V^\mu(x)=
\int{ d^2k_\perp dk^+ \theta(k^+)\over (2\pi)^{3/2}\sqrt{2k^+}}
\sum_{\omega=1,3}\bar\epsilon^\mu(\bbox{k},\omega)\left[
a(\bbox{k},\omega)e^{-ik\cdot x}
+a^\dagger(\bbox{k},\omega)e^{ik\cdot x}\right],\label{nvfield}
\end {equation} 
where the polarization vectors $\bar\epsilon^\mu(\bbox{k},\omega)$ are
given by\cite{yan34}:
\begin{eqnarray}
\bar\epsilon^\mu(\bbox{k},\omega)=
\epsilon^\mu(\bbox{k},\omega)-{k^\mu\over k^+}\epsilon^+(\bbox{k},\omega),
\end{eqnarray}
with the properties
\begin{eqnarray}
k^\mu\bar\epsilon_\mu(\bbox{k},\omega)=-{m_v^2\over k^+}
\epsilon^+(\bbox{k},\omega),
\quad \bar\epsilon_\mu(\bbox{k},\omega)
\bar\epsilon^\mu(\bbox{k},\omega')=-\delta_{\omega\omega'}+{m_v\over k^+}^2\;
\bar\epsilon^{\;+}(\bbox{k},\omega)\bar\epsilon^{\;+}(\bbox{k},\omega')\nonumber\\
\sum_{\omega=1,3}\bar\epsilon^\mu(\bbox{k},\omega)
\bar\epsilon^\nu(\bbox{k},\omega)=-(g^{\mu\nu}-g^{+\mu}{k^\nu\over k^+}
-g^{+\nu}{k^\mu\over k^+}).
\label{ebar}\end{eqnarray}

The path towards the light front 
Hamiltonian proceeds via the energy momentum tensor,
which is given by 
\begin{eqnarray}
T^{\mu\nu}=-g^{\mu\nu}{\cal L}\;+\;V^{\alpha\mu}V^{\nu\beta}g_{\beta\alpha}
+m_v^2V^\mu V^\nu
+{1\over 2}\bar\psi'[\gamma^\mu(i\partial^\nu-g_vV^\nu) 
+\gamma^\nu(i\partial^\mu-g_vV^\mu) ]\psi'\nonumber\\
+\partial^\mu\phi\partial^\nu\phi
+\partial^\mu\bbox{\pi}\cdot \partial^\nu \bbox{\pi}
+\bbox{\pi}\cdot\partial^\nu\bbox{\pi}
{\bbox{\pi}\cdot\partial^\mu\bbox{\pi}\over\pi^2}(1-{f^2\over \pi^2}
\mbox{sin}^2{\pi\over f}).
\end{eqnarray}
The use of the Fermion field equation allows one to obtain the 
light front Hamiltonian density
\begin{eqnarray}
T^{+-}=\bbox{\nabla}_\perp\phi\cdot\bbox{\nabla}_\perp\phi +m_\phi^2\phi^2
+{1\over 4}(V^{+-})^2+{1\over 2}V^{kl}V^{kl} +m^2_vV^kV^k\nonumber\\
+(\bbox{\nabla}_\perp\bbox{\pi})^2+ 
{\left({1\over 2}\bbox{\nabla}_\perp\pi^2\right)^2\over \pi^2}
\left(1-{f^2\over \pi^2}\mbox{sin}^2{\pi\over f}\right)+m_\pi^2f^2 
\mbox{sin}^2{\pi\over f}
\nonumber\\
+2\psi^\dagger_+\left(i {1\over 2}\stackrel{\leftrightarrow}{\partial}^-
-g_v\bar V^-\right)\psi_+.\label{tpm1}
\end{eqnarray}

It is now worthwhile to discuss a subtle feature regarding chiral symmetry in
light front formalisms.
Chiral invariance is defined as invariance under the transformation defined by
Eq.(\ref{chiral}) if the equal time formalism is used. Now the independent 
fermion variable is  $\psi_+$ and $\psi_-$ is a functional of this.
Thus chiral invariance is the invariance under the transformation
\begin{equation}
\psi_+\to e^{i\gamma_5\bbox{\tau}\cdot\bbox{a}}\psi_+\label{newchiral}
\end{equation}
which is not the same as Eq.(\ref{chiral}) because Eq.(\ref{newchiral})
produces a change in $\psi_-$ that is different than using 
$\psi_-\to e^{i\gamma_5\bbox{\tau}\cdot\bbox{a}}\psi_-$ \cite{mustaki,Wi 94}.
The $T^{+-}$ (or equivalently the light front Hamiltonian) 
of Eq.(\ref{tpm1}) is invariant under the transformation
(\ref{newchiral}) if the pion mass is neglected  so the usual chiral 
properties are obtained in these light front dynamics.

The expression (\ref{tpm1}) 
is useful for situations, such as in the mean field 
approximation case for infinite nuclear matter 
examined below, for which a simple expression
for 
$\psi_+$ is known. This is not always the case, so it is worthwhile
to use the Dirac equation to express 
$T^{+-}$ in an alternate form:
\begin{eqnarray}
T^{+-}=\bbox{\nabla}_\perp\phi\cdot\bbox{\nabla}_\perp\phi +m_\phi^2\phi^2
+{1\over 4}(V^{+-})^2+{1\over 2}V^{kl}V^{kl} +m^2_vV^kV^k\nonumber\\
+(\bbox{\nabla}_\perp\bbox{\pi})^2+ 
{\left({1\over 2}\bbox{\nabla}_\perp\pi^2\right)^2\over \pi^2}
\left(1-{f^2\over \pi^2}\mbox {sin}^2{\pi\over f}\right)+m_\pi^2f^2 
\mbox{sin}^2{\pi\over f}
\nonumber\\
+\bar\psi\left(\bbox{\gamma}_\perp
\cdot(\bbox{p}_\perp-g_v\bbox{\bar V}_\perp
)+(MU+g_s\phi)\right)\psi.\label{tpm2}
\end{eqnarray}

It is convenient to consider $\psi_-$ as a sum of terms, one $\xi_-$
whose relation with $\psi_+$ is free of interactions\cite{des71}, 
 the other $\eta_-$ containing the interactions. That is, rewrite the second
of Eq.~(\ref{yan}) as \cite{harizhang}
\begin{eqnarray}
\xi_-&=&{1\over p^+}(\bbox{\alpha}_\perp\cdot 
\bbox{p}_\perp+\beta M)\psi_+\nonumber\\
\eta_-&=&{1\over p^+}(-\bbox{\alpha}_\perp\cdot 
g_v\bbox{\bar V}_\perp+\beta(M(U-1)+g_s\phi))\psi_+. \label{yan1}
\end{eqnarray}
Furthermore, define $\xi_+(x)\equiv\psi_+(x)$, so that 
\begin{equation}
\psi(x)=\xi(x)+ \eta_-(x), \label{fcon}
\end{equation}
where $\xi(x)\equiv \xi_-(x)+\xi_+(x)$.
The purpose of the above decomposition is to separate the dependent and
independent parts of $\psi$ and to allow one to  expand $\xi$ in terms of
eigenstates of momentum. 

One may make a similar treatment for the vector meson fields. The operator
$V^{+-}$, determined by Eq~(\ref{vpm}), is relevant for the Hamiltonian.
Part of this operator is determined by a constraint equation. To see this
examine  Eq~(\ref{vpm}), and make a definition:
\begin{equation}
V^{+-}=v^{+-}+\omega^{+-}, \label{vcon}
\end{equation}
where 
\begin{equation}\omega^{+-}={-2\over \partial^+}J^+.
\end{equation}
Next use equations (\ref{fcon}) and (\ref{vcon}) to rewrite the
Hamiltonian as a sum of a free and interacting terms. 
The sum of the last term of Eq\.~(\ref{tpm2}) and the terms
involving $\omega^{+-}$ is the density of the  interaction 
Hamiltonian, $P_I^-$, plus the free fermion term, $P^-_0(N)$. 
Use equations (\ref{fcon}) and (\ref{vcon}) in  the expression (\ref{tpm2})
for $T^{+-}$ along with the 
field equations and integration by parts to find:
\begin{equation}
P_0^-(N)={1\over 2}\int d^2x_\perp dx^-
\bar\xi\left(\bbox{\gamma}_\perp\cdot\bbox{p} +M\right)\xi, \label{freef}
\end{equation}
and
\begin{equation}
P^-_I=v_1+v_2+v_3,  \label{defv}
\end{equation}
with 
\begin{equation}
v_1=\int d^2x_\perp dx^-\bar\xi\left(g_v\gamma\cdot\bar V+M(U-1)+g_s\phi
\right)\xi,\label{v1}
\end{equation}
\begin{equation}
v_2=\int d^2x_\perp dx^-\bar\xi\left(-g_v\gamma\cdot\bar V
+M(U-1)+g_s\phi\right)\;{\gamma^+\over 2p^+}\;\left(-g_v\gamma\cdot\bar V
+M(U-1)+g_s\phi
\right)\xi, \label{v2}
\end{equation}
and 
\begin{equation}
v_3={g_v^2\over32}\int d^2x_\perp dx^-\int dy^-_1
\bar\xi(\bbox{x}_\perp,y^-_1)\gamma^+\xi(\bbox{x}_\perp,y^-_1)
\epsilon(x^--y^-_1)\int dy^-_2\epsilon(x^--y^-_2)
\bar\xi(\bbox{x}_\perp,y^-_2)\gamma^+\xi(\bbox{x}_\perp,y^-_2).\label{v3}
\end{equation}

The term $v_1$ accounts the emission or absorption of a
 single vector or scalar meson, as well as the emission or absorption
of any number of pions through the operator $U-1$. The term $v_2$ 
includes contact terms in which there is propagation of an instantaneous 
fermion. The term $v_3$ accounts for the propagation of an instantaneous
vector meson.

We may now quantize the 
fermion  fields using
\begin{equation}
\xi(x)=\int{ d^2k_\perp dk^+ \theta(k^+)\over (2\pi)^{3/2}\sqrt{2k^+}}
\sum_{\lambda=+,-}\left[u(\bbox{k},\lambda)e^{-ik\cdot x}b(\bbox{k},\lambda)+
v(\bbox{k},\lambda)e^{+ik\cdot x}d^\dagger(\bbox{k},\lambda)\right],
\label{dq}
\end{equation}
where again the momenta are on shell: $k^-={\bbox{k}_\perp^2 +M^2\over k^+}$,
and the anti-commutation relations are given by 
\begin{eqnarray}
\{b(\bbox{k},\lambda),b^\dagger(\bbox{k}',\lambda')\}=
\{d(\bbox{k},\lambda),d^\dagger(\bbox{k}',\lambda')\}=\delta_{\lambda,\lambda'}
\delta^{(2,+)}(\bbox{k}-\bbox{k}'),\nonumber\\ 
\{b(\bbox{k},\lambda),b(\bbox{k}',\lambda')\}=
\{d(\bbox{k},\lambda),d(\bbox{k},\lambda')\}=0. \label{fcomm}
\end{eqnarray}
The properties of the Dirac spinors are described in Appendix 
A. The term $P^-_0(N)$ of Eq.~(\ref{freef})
can now  be expressed as 
\begin{equation}
P^-_0(N)=\int d^2k_\perp dk^+\theta(k^+){k_\perp^2+M^2\over k^+}
\sum_\lambda(b^\dagger(\bbox{k},\lambda)b(\bbox{k},\lambda)
+ d^\dagger(\bbox{k},\lambda)d(\bbox{k},\lambda)).
\end{equation}

The component that is related to the plus momentum is $T^{++}$.
The necessary expression is given by 
\begin{eqnarray}
T^{++}=V^{ik}V^{ik}
+m_v^2V^+ V^+
+\bar\psi\gamma^+ i\partial^+ \psi\nonumber\\
+\partial^+\phi\partial^+\phi
+\partial^+\bbox{\pi}\cdot \partial^+ \bbox{\pi}
+\bbox{\pi}\cdot\partial^+\bbox{\pi}
{\bbox{\pi}\cdot\partial^+\bbox{\pi}\over\pi^2}(1-{f^2\over \pi^2}
\mbox{sin}^2{\pi\over f}). \label{tpp}
\end{eqnarray}

\section{Chiral Symmetry and Pion-Nucleon Scattering}
We begin by  showing that, if one starts with a non-linear representation of
chiral symmetry,  the requirement of solving the constraint 
equation for the $-$ component of the fermion field leads one to a Lagrangian
of the Gursey-type  linear representation. 

The focus is on chiral properties and pion-nucleon scattering, so we dispense
with the vector and non-chiral $\phi$ meson fields for this section, and
it is sufficient
to examine  only the following fermion-pion term
 of a non-linear representation\cite{bira}:
\begin{equation}
{\cal L}_{N\pi}=\bar{N}\left[\gamma_\mu i\partial^\mu-M+{1\over 1+(\pi/2f)^2}
\left({1\over 2f}\gamma^\mu\gamma_5\bbox{\tau\cdot}\partial^\mu\bbox{\pi}-
({1\over 2f})^2
\gamma^\mu\bbox{\tau\cdot \pi\times}\partial^\mu\bbox{\pi}\right)\right]N.
\end{equation}
Next obtain the fermion field equation and
make the usual decomposition:  $N_\pm\equiv
\Lambda_\pm N$ with
\begin{eqnarray}
\left(i\partial^--O^-\right)N_+=
[\bbox{\alpha}_\perp\cdot(\bbox{p}_\perp-\bbox{O}_\perp)
+\beta M
]N_-\label{nn}
\nonumber\\
\left(i\partial^+-O^+\right)N_-=[
\bbox{\alpha}_\perp\cdot\bbox{(p}_\perp-\bbox{O}_\perp)+\beta M]N_+,
\end{eqnarray}
where the operator $O^\mu$ has been defined as 
\begin{equation}
O^\mu\equiv {-1\over 1+(\pi/2f)^2}
\left({1\over 2f}\gamma_5\bbox{\tau\cdot}\partial^\mu\bbox{\pi}-
({1\over 2f})^2
\bbox{\tau\cdot \pi\times}\partial^\mu\bbox{\pi}\right).
\end{equation}
We wish to remove the $O^+$ term from the left hand side of the equation
for $N_-$. This can be done by defining a unitary operator $F$ and fermion
field $\chi$ such
that 
\begin{equation}N=F \chi\label{defchi}\end{equation}
 with \begin{equation}i\partial^+F=O^+F.\label{feq}\end{equation}
 The identity\cite{gursey} 
\begin{equation}
U_2^{{1\over2}}\partial^\mu U_2^{-{1\over2}}=i O^\mu,
\end{equation}
where $U_2$ is given in Eq.~(\ref{us}),
helps a good deal. Its use in Eq.~(\ref{feq}), combined with the 
condition $\partial^\mu (U_2 U_2^{-1})=0$, leads to the result
\begin{equation}
F=U_2^{{1\over2}} \label{ff},
\end{equation}
so that 
using Eqs. (\ref{ff}) and (\ref{defchi}) in (\ref{nn}) yields
\begin{eqnarray}
i\partial^-\chi_+=\left[
\bbox{\alpha}_\perp\cdot \bbox{p}_\perp+\beta MU_2\right]\chi_-
\nonumber\\
i\partial^+\chi_-=\left[
\bbox{\alpha}_\perp\cdot\bbox {p}_\perp+\beta MU_2\right]\chi_+.
\end{eqnarray}
This is of the desired form in which no interactions appear on the 
left-hand-side of the equation for $\chi_-$.
Thus the use of light front quantization mandates that the 
pion-nucleon interactions be of the form of Eq.~(\ref{lag}).

The first test for any chiral formalism is to reproduce the early soft pion
theorems\cite{softpi}.
Here we concentrate on low energy pion nucleon scattering
because of its relation to the nucleon-nucleon force. We work to second 
order in $1/f$ in this first application. In this case, each of the $U_i$
takes the same form:
\begin{equation}
U=1+i\gamma_5 {\bbox{\tau\cdot\pi}\over f} -{1\over 2f^2}\pi^2. \label{um1}
\end{equation}

This expression is to be used in the potentials 
$v_1$ and $v_2$ of Eqs.~(\ref{v1}) and (\ref{v2}). The second order scattering
graphs are of three types and are shown as time $x^+$ ordered 
perturbation theory diagrams in Fig.~1. The kinematics are such that
$\pi (q) N(k)\to \pi(q') N(k')$, with $P_i=q+k$ and $P_f=q'+k'$.
 The iteration
of $v_1$ to second order yields the direct and crossed graphs of Fig.~1a. 
In this formalism  $v_1$ is proportional to the matrix element of $\gamma_5$ 
between $u$ spinors, so it is  
proportional to the momentum of the absorbed or  
emitted pion.
 Thus the terms of Fig.~1a vanish near threshold. The terms of Fig.~1b are
generated by the $\bar u \gamma_5 v$ terms of $v_1$. Using the  various field 
expansions in the expression (\ref{v1}) for $v_1$ leads to the result that 
plus-momentum is conserved and the plus momentum of 
every  particle is greater than zero. This means that the first of Fig.1b 
vanishes identically and the second vanishes for values of the initial 
pion plus momentum that are less than twice the nucleon mass.
 The net result is that only the instantaneous term of $v_2$ and 
the $\pi^2$ 
term of $v_1$ (shown in Fig. 1c) remain to be evaluated.

Proceeding more formally, we evaluate the S-matrix given by
\begin{equation}
S=T_+e^{-{i\over2}\int^\infty_{-\infty} dx^+ \hat P_I^-(x^+)} \label{smat1}
\end{equation}
where $T_+$ is the $x^+$ (light-front time) ordering operator and
$\hat P_I^-$ is the interaction representation light front Hamiltonian. Then
\begin{equation}
(S-1)_{fi}=-2\pi i\delta (P_i^--P_f^-)\langle f|T(P^-_i)|i\rangle,\label{smat2}
\end{equation}
with 
\begin{equation}
T(P^-_i)= P^-_I+ P^-_I{1\over P_i^--P^-_0}T(P^-_i)
\end{equation}
The evaluation proceeds by using the field expansions in the expressions
for $v_1$ and $v_2$. Integrating over $d^2x_\perp dx^+$ and evaluating the
result between the relevant initial and final pion-nucleon states
leads to the result that each contribution to the S-matrix
 is proportional to a common factor,
$${\delta^{(2,\perp)}(P_i-P_f)\over 2 (2\pi)^3 \sqrt{k'^+k^+q'^+q^+}},$$
which combines with the result of the required integration over the light cone
time ($x^+$) to provide the necessary momentum conservation and flux factors.
The remaining factor of each term is its 
contribution to the invariant amplitude ${\cal M}$. The result is
\begin{equation}
{\cal M}=\tau_i\tau_f {M^2\over f^2} {\bar u(k')\gamma^+u(k)\over 2(k^++q^+)} +
\tau_f\tau_i {M^2\over f^2} {\bar u(k')\gamma^+u(k)\over 2(k^+-q^+)} 
-\delta_{if}{M\over f^2}\bar u(k')u(k)
\end{equation}
where the three terms here correspond to the three terms of Fig.~1c. The role
 of cancellations in the reduction of  the term proportional to $\delta_{if}$ 
is already apparent.
To understand the threshold physics take $k'^+=k^+=M$ and $q'^+=q^+=m_\pi$.
Then one finds 
\begin{equation}
{\cal M}= 
\delta_{if}{2m_\pi^2\over f^2} +2i\epsilon_{fin}\tau_n{m_\pi M\over f^2}
\end{equation}
to leading order in $m_\pi/M$. The weak nature 
of the $\delta_{if}$ term and the presence of the second Weinberg-Tomazowa term
is the hallmark of chiral symmetry\cite{softpi}.

The same results could be obtained using the linear sigma model, with
$\sigma$ exchange playing the role of  the $\pi^2$ term of 
Eq.(\ref{um1}). 

\section{Nucleon Nucleon Scattering via One-Boson Exchange Potentials}
The ultimate aim is to derive the nuclear wave function including 
 correlation effects. The first step is to understand
 nucleon-nucleon scattering using our light front formalism.
We start with  the one boson exchange approximation, 
discuss the light front wave equation and 
show that this procedure gives the
same scattering amplitude as the usual procedure of computing the
one-boson exchange contributions to the invariant amplitudes and using
the Blankenbeckler-Sugar reduction of
the Bethe-Salpeter equation\cite{rm,rm1}. This usual
procedure is covariant, so that our construction shows that 
 the light front wave procedure respects rotational invariance.
This invariance is 
the result of Strikman and Frankfurt\cite{fs1} and others.
The present treatment
explicitly includes the effects of nucleon spin and the 
nucleon-nucleon interaction can  derived from an underlying chiral Lagrangian.

The starting point is the S matrix of Eqs.~(\ref{smat1}) and (\ref{smat2}).
Here the initial state $i$ consists of nucleons with quantum numbers labeled
by 1 and 2, and the state $f$ consists of nucleons 3 and 4. To be definite,
we take the plus momentum  of nucleon 1 to be greater than that of 
nucleon 3, and the momentum transfer $q$ to be 
\begin{equation}q\equiv k_1-k_3,\end{equation}
 so that $q^+>0$.

The lowest order contributions to the invariant amplitude are represented
by the light-front-time ordered graphs shown in Fig.~2. The graphs of Fig.~2a
represent terms of the form $v_1 {1\over P_i^--P_0^-}v_1$, and that of
Fig. ~2b accounts for the instantaneous massive vector boson exchange term
of $v_3$. These terms may be evaluated by using the field expansions and
doing the relevant integrals over the $d^2x_\perp dx^-$ coordinate space.
Each term has a common factor of 
$${4M^2\delta^{(2,\perp)}(P_i-P_f)\over 2\sqrt{k_1^+ k_2^+ k_3^+ k_4^+}},$$
 where the 
factor $4M^2$ in the numerator is compensated by dividing the invariant 
amplitude by $4M^2$. 

It is simplest to consider the effects of  scalar $\phi$
and pseudoscalar $\bbox\pi$ exchanges
at the same time. The scattering amplitudes $\langle 3,4|
\cal K(\phi,\bbox{\pi})|1,2\rangle$
 take the form
\begin{equation}
\langle 3,4|{\cal K}(\phi,\bbox{\pi}) |1,2\rangle
= {\bar u(4)\Gamma u(2) \;\bar u(3)\Gamma u(1) \over
4M^2 (2\pi)^3 k^+\left(k_1^--k_3^--k^-\right)}, \label{gen}
\end{equation}
where the notation is that $u(i)$ is the spinor 
for a nucleon of quantum numbers $i$, and $\Gamma$ is either of the
form $g_s$  or $i\;g\gamma_5$. The momentum of the exchanged meson is $k$, and
it is necessary to realize that
\begin{eqnarray}
k^+=q^+,\;\bbox{k}_\perp=\bbox{q}_\perp \label{off}
\end{eqnarray}
but 
\begin{eqnarray}
k^-={k_\perp^2+\mu^2\over k^+}\ne q^-, \label{off1}
\end{eqnarray}
where $\mu$ is the mass of the exchanged scalar meson or pion.
The factor $1/k^+$ arises from the denominators of the field expansions
and $\left(k_1^--k_3^--k^-\right)$ is the result of evaluating
the light front energy denominator $ P_i^--P_0^-$. Define the energy 
denominator of eq.(\ref{gen}) to be $D$ so that
\begin{eqnarray}
D=k^+\left(k_1^--k_3^--k^-\right)= (k^+_1-k^+_3)(k_1^--k_3^-)-k^+k^-.
\end{eqnarray}
Using Eqs.~(\ref{off}) and (\ref{off1}) immediately yields
\begin{eqnarray}D=q^2-\mu^2,
\end{eqnarray}
so the amplitude  $\cal K$ takes the form
\begin{equation}
\langle 3,4|{\cal K}(\phi,\pi)|1,2\rangle
 = {\bar u(4)\Gamma u(2)\; \bar u(3)\Gamma u(1) \over
4M^2 (2\pi)^3 \left(q^2-\mu^2\right)}.
\end{equation}
This is the usual \cite{rm,rm1,geb} expression for a one-boson exchange 
potential, if no form 
factor effects are included. Note that the Klein-Gordon propagator is obtained
using  only a single time-ordered graph. The calculation with  the 
equal-time formulation
requires the summation of two time-ordered graphs.

The derivation of the contribution of vector meson exchange proceeds by 
adding the terms of Fig 2a and 2b. The term of Fig.~2a can immediately seen to
be 
\begin{equation}
\langle 3,4|{\cal K}_{2a}(V)|1,2\rangle = g_v^2
{\bar u(4)\gamma_\mu u(2) \;\bar u(3)\gamma_\nu u(1) \over
4M^2 (2\pi)^3 \left(q^2-m_v^2\right)} 
\left[-g^{\mu\nu}+g^{+\mu\;}{k^\nu\over k^+}+g^{+\nu\;}{k^\mu\over k^+}\right].
\end{equation}
The factor in brackets arises from 
the polarization sum, recall Eq. (\ref{ebar}). 
It is worthwhile to define the contribution  of the second two terms in the
bracket, which
result from the difference between $\bar V^\mu$ and $V^\mu$
 fields,   as $\langle 3,4|{\cal K}_{bar}|1,2\rangle$, with
\begin{equation}
\langle 3,4|{\cal K}_{bar}|1,2\rangle
=g_v^2 {\bar u(4)\gamma^+u(2) \;\bar u(3)\gamma\cdot k u(1) 
+\bar u(4)\gamma\cdot k u(2) \;\bar u(3)\gamma^+u(1)\over
4M^2 (2\pi)^3 q^+\left(q^2-m_v^2\right)}.\label{bar}
\end{equation}
Next use the relations 
$\bar u(3)\gamma\cdot q\; u(1)=\bar u(4)\gamma\cdot q\; u(2)=0 $ and the 
equality of the $+$ and $\perp$ components of  $k$ with those of $q$ to 
obtain the results
\begin{eqnarray}\bar u(3)\gamma\cdot k\; u(1)={1\over 2}\bar u(3)\gamma^+u(1)
(k^--q^-),\;\nonumber\\
\bar u(4)\gamma\cdot k u(2)={1\over 2}\bar u(4)\gamma^+u(2)
 (k^--q^-).
\end{eqnarray}
But  $k^--q^-={q_\perp^2+m_v^2\over q^+}-q^-=-{q^2-m_v^2\over k^+}$, 
which leads to a compact rewriting of Eq.~(\ref{bar}) as 
\begin{equation}
\langle 3,4|
{\cal K}_{bar}|1,2\rangle
=-g_v^2 {\bar u(4)\gamma^+u(2) \;\bar u(3)\gamma^+ u(1) 
\over
4M^2 (2\pi)^3 (k^+)^2}
\end{equation}
The term of Fig.~2b is obtained
by using the field expansion in the equation for $v_3$, (\ref{v3}), integrating
over coordinate space and removing the common factor. The result is
\begin{equation}
\langle 3,4|{\cal K}_{2b}(V) |1,2\rangle
= g_v^2{\bar u(4)\gamma^+ u(2) \;\bar u(3)\gamma^+ u(1) \over
4M^2 (2\pi)^3 (k^+)^2 }, 
\end{equation}
which exactly cancels the term $\langle 3,4|{\cal K}_{bar}|1,2\rangle$. 
The net result is that
the amplitude for vector meson exchange,
 $\langle 3,4|{\cal K}(V)|1,2\rangle
=\langle 3,4|{\cal K}_{2a}(V)+{\cal K}_{2b}|1,2\rangle$,
takes the familiar form:
  \begin{equation}
\langle 3,4|{\cal K}(V)|1,2\rangle = -g_v^2
{\bar u(4)\gamma_\mu u(2) \bar u(3)\gamma^\mu u(1) \over
4M^2 (2\pi)^3 \left(q^2-m_v^2\right)}.
\end{equation}

The sum
of the amplitudes arising from each of the 
individual one boson exchange terms:
\begin{equation}
\langle 3,4|{\cal K}|1,2\rangle=\langle 3,4|
{\cal K}(\phi)+{\cal K}(\bbox {\pi})+{\cal K}(V)|1,2\rangle, 
\end{equation}
gives  the invariant amplitude to second order in each of the
coupling constants.

These amplitudes are strong, so computing the nucleon-nucleon scattering 
amplitude and phase shifts requires including higher order terms.
One may include  a sum which gives unitarity by including 
all iterations of the scattering operator ${\cal K}$ 
through intermediate two-nucleon states:
\begin{equation}
{\cal M}={\cal K} +{\cal K} {P_{2N}\over P_i^--P^-_0 }{\cal M},\label{iter}
\end{equation}
where $P_i^-$ is the negative-momentum in the initial state and 
$P_{2N}$ projects on to two-nucleon intermediate states.
More explicitly, Eq.~(\ref{iter}) is  given by
\begin{equation} 
\langle3,4|{\cal M}|1,2\rangle
=\langle 3,4|{\cal K}|1,2\rangle+
\sum_{\lambda_5,\lambda_6}\int  \langle 3,4|{\cal K}|5,6\rangle 
{2M^2\over p_5^+p_6^+}
{d^2p_{5\perp}dp^+_5\over P_i^--(p_5^-+p_6^-)+i\epsilon}
\langle5,6|{\cal M}|1,2\rangle.
\end{equation} 
after removing the common factor and accounting for the 
momentum conserving delta
functions. One realizes that this
is of the form of the Weinberg equation\cite{We66}
by expressing the plus-momentum   variable in terms of a light-front 
momentum  fraction
$\alpha$ such that
\begin{equation}
p_5^+=\alpha P_i^+,
\end{equation}
and using the relative and total momentum variables:
\begin{eqnarray}
\bbox{p}_\perp\equiv (1-\alpha)\bbox{p_5}_\perp-\alpha
\bbox{p_6}_\perp\nonumber\\
\bbox{P_i}_\perp=\bbox{p_5}_\perp+\bbox{p_6}_\perp
\end{eqnarray}
Then
\begin{equation} 
\langle3,4|{\cal M}|1,2\rangle
=\langle 3,4|{\cal K}|1,2\rangle+
\int\sum_{\lambda_5,\lambda_6} \langle 3,4|{\cal K}|5,6\rangle 
{2M^2\over \alpha(1-\alpha)}
{d^2p_\perp d\alpha\over P_i^2-{p_{\perp}^2+M^2\over\alpha(1-\alpha)}
+i\epsilon}
\langle5,6|{\cal M}|1,2\rangle, \label{419}
\end{equation}
where $P_i^2$ is square of the total initial  four-momentum,
otherwise known as the invariant energy $s$ and 
${p_{\perp}^2+M^2\over\alpha(1-\alpha)}$
 is the corresponding quantity for the intermediate state.
Because the kernal ${\cal K}$ is itself an invariant amplitude
the procedure of solving this equation to  determine observables is
manifestly 
covariant. 

Equation~(\ref{419}) can
in turn can be re-expressed as the Blankenbecler-Sugar (BbS) equation
\cite{BbS} 
by  using the variable transformation\cite{Te 76}:
\begin{equation}
\alpha={E(p)+p^3\over 2E(p)}, \label{alpha}
\end{equation}
with $E(p)\equiv\sqrt{\bbox{p}\cdot\bbox{p}+M^2}$.
The result is:
\begin{equation}
\langle3,4|{\cal M}|1,2\rangle
=\langle3,4|{\cal K}|1,2\rangle+\int
\sum_{\lambda_5,\lambda_6} \langle 3,4|{\cal K}|5,6\rangle
{M\over E(p)}
{d^3p \over {p_i^2-p^2\over M}
+i\epsilon}
\langle5,6|{\cal M}|1,2\rangle, \label{bsbs}
\end{equation}
which is the desired equation. 
The three-dimensional propagator is exactly that of the BbS equation; there is
one difference. Our one boson exchange potentials depend on the 
square of the four momentum $q^2$ transferred when a meson is absorbed or
emitted by a nucleon. Thus the energy difference between the initial and final
on-shell nucleons is included and $q^0\ne 0$.
 The derivation of the BbS equation from
the Bethe-Salpeter equation specifies that  $q^0=0$ is used 
in the meson propagator. Including $q\ne 0$ instead of $q^0=0$ increases the
range of the potential. Such an effect can be hidden in phenomenological
potentials by changing the pion-nucleon coupling constant or form factor.

One can easily convert Eq.~(\ref{bsbs}) into
the Lippman-Schwinger equation of non-relativistic scattering theory
by removing the factor $M/E(p)$ with a simple transformation\cite{pl}.

\subsection{Comparison with Realistic One-Boson Exchange Potentials}
The present results are that one can use the light front technique to
derive nucleon-nucleon potentials in the one-boson exchange OBE
approximation and use these in an appropriate wave equation. Therefore
our procedure is directly comparable to the one used in
constructing the realistic Bonn  one-boson exchange potentials used in
momentum space. 
Those potentials also have a close connection with an underlying
Lagrangian. Our purpose here is to argue that the present procedure
can yield potentials essentially identical to the Bonn OBEP potentials and
therefore would lead to a good description of the NN data.

The Bonn one-boson exchange  potentials employ six different mesons
$\pi,\eta, \omega,\rho, \sigma$ and the (isovector scalar)
$\delta$ meson. The present techniques can be used to handle all of these
mesons and their couplings, with the possible exception of the tensor 
$\sigma_{\mu\nu}q^\nu$ part of the $\rho$-nucleon interaction. 

The presence of such a tensor interaction makes it difficult (or
impossible) to write the equation for $\psi_-$ as $\psi_-=1/p^+\cdots
\psi_+.$ This is relevant because the standard value of the ratio of
the tensor to vector $\rho$-nucleon coupling $f_\rho/g_\rho$ is 6.1,
based upon Ref.~\cite{hp} and subsequent papers. Reproducing the
observed values of $\varepsilon_1$ and P-wave wave phase shifts
requires a large value 
$f_\rho/g_\rho$; see Ref.~\cite{brm}.  However the Lagrangian
compensates for its lack of a $\rho$-N interaction with tensor
coupling by generating such a term via vertex correction diagrams
(which are the origin of the anomalous magnetic moment of the electron
in QED). Such diagrams probably do not generate the
phenomenologically required values of the coupling constants, but all
that is needed here is that terms of the correct form be
produced. This is because the standard procedure is to choose the
values of the coupling constants so as to yield a good description of
the NN scattering data.  Indeed the potentials A,B, and C are defined
by the parameters which account for the mesonic masses, coupling
constants and form factors. Thus we end up with the same procedure that is
used in the Bonn one boson exchange potentials.

This brings us to the treatment of divergent terms in our procedure.
The definition of any effective Lagrangian
requires the specification of such a procedure.  For the present, 
it is sufficient to say that we introduce form factors, $F_\alpha(q^2)$
which reduce the strength of the $\alpha$ meson-nucleon coupling for
large values of $-q^2$. This is also the procedure of Ref~\cite{rm,rm1}.


The net result is that the one-boson exchange treatment of the
nucleon-nucleon potential and the T-matrix resulting from its use in
the BbS equation is essentially the same as the one-boson exchange
procedure of Ref.~\cite{rm,rm1}. Thus our light front treatment is
guaranteed to be consistent with nucleon-nucleon nucleon scattering
data measured in the standard energy range. Such a similarity has also
been obtained using relativistic Hamiltonian dynamics\cite{Fu 91}.

\subsection {Nucleonic Contribution to the Two Pion Exchange Potential}
The dominant contribution to the two pion exchange potential arises
from contributions to intermediate states that include one or two
$\Delta$'s\cite{rm}, and a treatment of such effects
based on chiral symmetry has been provided by 
van Kolck and collaborators\cite{biraref}. 

Including the effects of 
$\Delta$'s is beyond the scope of the present 
work, but we are able to 
discuss the  two pion exchange contribution (of order $(M/f)^4$)
to the nucleon nucleon
potential.  The property that a sum of light cone
time-ordered diagrams is equal a single Feynman graph can be used to
simplify the calculation. The relevant Feynman graphs are displayed in
Fig.~3; the terms originating from the linear
$\gamma_5{\bf\tau}\cdot{\bf \pi}$ coupling (a,b), from the quadratic
$\pi^2-N$ coupling (c) and from a combination of the linear and
quadratic interactions (d) are indicated.  The line through the
two-nucleon intermediate state of Fig.~3a is meant to indicate that
the contribution arising from iterating the one pion exchange
interaction is removed. This has been a standard procedure since the
work of Ref.~\cite{pl}, and will not be discussed further.

The sum of the terms of Fig.~3a and 3b is equal to the Partovi-Lomon
two pion exchange potential, as they  used the pseudoscalar pion-nucleon
interaction. This interaction certainly simplifies the calculation; in
particular the diagrams of Fig.~3a, b and d are convergent (whereas
they would be strongly divergent if pseudovector coupling were to be
used.  One can use such a pseudoscalar coupling, and include the
effects of chiral symmetry, provided one also includes the effects of
the $\pi^2-N$ coupling shown in Fig.~3c, and the combined effects of
the linear and quadratic interactions, Fig.~3d.  The quadratic
interaction term cancels the large pair terms in pion-nucleon
scattering and should also play a significant role here in reducing
the size of the computed potential.  Thus we expect that the
Partovi-Lomon potential contains too large an attraction. 

Next turn to the procedure used in constructing the full Bonn
potential.  This potential is constructed by ignoring all of the
Z-graphs and including the the effects of the two-nucleon intermediate
states which arise from the crossed graph, Fig.~3b, as well as the
parts of Fig.~3a arising from time ordered terms in which two pions
exist at the same time. (For such contributions to the TPEP the linear
pseudoscalar and pseudovector interactions are are evaluated between
on shell positive energy nucleon spinors, and are therefore
equivalent.  The resulting contribution to the TPEP is small, but is
comparable to that of the iterated OPEP.  The neglect of the Z graphs
goes a long way towards including the effects of chiral
symmetry. However, terms involving the Weinberg-Tomazowa interaction
at one or two vertices are ignored. The computation of the graphs of
Fig.~3 would include such effects implicitly as well as that of pair
suppression. Thus a detailed comparison would be useful. However, 
the small nature of the effects that we discuss now indicate that the 
dominance of the 
TPEP by effects of intermediate
$\Delta$'s will remain unchallenged.

\section{Mean Field Approximation}
The nucleon-nucleon interaction of the previous section can be used as
the basis for a light front Brueckner theory of nuclei. We 
study the mean field approximation for infinite nuclear matter
as a first step.  The nuclear mean field
model- the shell model- occupies pre-eminence in  understanding
nuclear structure. We need to
see if our formalism can describe this physics.

In the mean field approximation\cite{bsjdw}, the coupling
constants are  
considered strong and the Fermion density  large. Then the meson 
fields can be approximated as classical-  the 
sources of the meson fields  are replaced
 by their expectation values. In this case, the nucleon mode functions will
be plane waves and the nuclear matter
ground state can be 
 assumed to be a  normal Fermi gas,  of Fermi momentum
$k_F$, and  of large volume $\Omega$ in its rest frame. 
We consider the case that there is an equal  number of protons and neutrons.

First we 
examine  the mesonic field equations (\ref{vmeson})-(\ref{pimeson}). The baryon
source of
the pion field is a pseudoscalar operator, so  its expectation value vanishes in
the ground  state. Thus this mean field
approximation leads to the result that $\pi_i\to 0$.
The other
 meson fields are constants, independent of space and time,  given by 
\begin{eqnarray}
\phi=-{g_s\over m_s^2} \langle \bar \psi \psi\rangle\label{smf}\\
V^\mu={g_v\over m_v^2} \langle \bar \psi
\gamma^\mu\psi\rangle=\delta^{0,\mu}{g_v\rho_B\over m_v^2},\label{mfa}
\end{eqnarray}
where the brackets denote  expectation values of the nuclear 
ground state in its rest frame and the baryon density is 
\begin{equation}\rho_B=2k_F^3/3\pi^2.\end{equation}
This result that $V^\mu$ is a constant, along with Eqs.~(\ref{vbar}) and 
(\ref{mfa}),  
can be used to determine $\bar V^\mu$. In particular, $\bar V^+=0$ by
construction. Furthermore, the conditions that $V^i=0$ and 
$\partial^i V^+=\partial^i V^0=0$ tell us that $\bar V^i=0$. Finally
$\partial^-V^+=0$, so that $\partial^+ \bar V^-=\partial^+ V^0$, so the net
result is that the 
only non-vanishing component of $\bar V^\mu$ is $\bar {V}^-=
V^0$.  

With this mean field approximation, the fermionic field
equations (\ref{yan}) can be rewritten as
\begin{eqnarray}
(i\partial^--g_v \bar V^-)\psi_+=(\bbox{\alpha}_\perp\cdot 
\bbox{p}_\perp+\beta(MU+g_s\phi))\psi_-\nonumber\\
i\partial^+\psi_-=(\bbox{\alpha}_\perp\cdot 
\bbox{p}_\perp +\beta(MU+g_s\phi))\psi_+. \label{nyan}
\end{eqnarray}
Now $\phi$ and $\bar V^-$ are constants so we expect the mode functions
for the field expansion of $\psi$ to be of the plane wave form
$\sim e^{ik\cdot x}$
and can be obtained from Eq.~(\ref{nyan}) as\cite{exp}
\begin{equation}
(i\partial^--g_v \bar V^-)\psi_+
={\bbox {k}_\perp^2 +(M+g_s\phi)^2\over k^+}\psi_+. \label{sol}
\end{equation}
The light front eigenenergy $(i\partial^-\equiv k^-)$
is the sum of a kinetic energy term in which the mass is shifted by the
presence of the scalar field, and an energy arising from the vector field.
Comparing 
this equation with the one for free nucleons, 
$k^-={k_\perp^2+M^2\over k^+}$, shows that   the nucleons 
have a  mass $M+g_s\phi$ and  move
in plane wave states. The nucleon 
field operator is constructed using the solutions of 
Eq.~(\ref{sol}) as the plane wave basis states. This means that 
the nuclear matter ground state, defined by operators that create and 
destroy baryons in eigenstates of Eq.~(\ref{sol}), is the correct 
wave function and that Equations~ (\ref{mfa}) and (\ref{sol})
represent the solution of the approximate
field equations, and the diagonalization of the  Hamiltonian. 

One question remains. We are going to fill up a Fermi sea, but $k_F$
is the magnitude of a three vector. How is this three vector defined?
This was answered in the paper of Glazek and Shakin \cite{gs} who
showed that rotational invariance is manifest if one uses the
definition:
\begin{equation}
k^+=\sqrt{(M+g_s\phi)^2+{\bf k}\cdot{\bf k}}+k^3,\label{kplus}
\end{equation}
which implicitly defines $k^3$.  Using Eq.~(\ref{kplus}) allows one to
maintain the equivalence between energies computed in the light front
and equal time formulations of scalar field theories\cite{bg}.  A
similar equation has been used to restore manifest rotational
invariance in light-front QED\cite{mp}. We shall show that this same
expression also restores rotational invariance in this mean field
problem when vector mesons are included.

Equation~(\ref{kplus}) has the correct form 
in the limit of  non-interacting nucleons and
therefore seems natural\cite{mini}.
We attempt a heuristic derivation of this equation
using the requirement that manifest rotational invariance be restored.
The starting point is the observation that 
Eq.~(\ref{alpha}), with its definition of $\alpha$ as the 
plus momentum fraction carried by a nucleon,
restores manifest rotational invariance in the 
two-nucleon system. Let's consider the mean field approximation as 
involving an interaction
between a nucleon and a very heavy particle containing A-1 nucleons 
(with $A\to \infty$). Then the variable $\alpha_A$, 
which is the fraction of the 
nuclear plus-momentum carried by a nucleon, is given by 
\begin{equation}
\alpha_A={\sqrt{\bbox {k}\cdot\bbox {k} +(M+g_s\phi)^2} +k^3\over
\sqrt{\bbox {k}\cdot\bbox{k} +M_{A-1}^2} + 
\sqrt{\bbox{k}\cdot\bbox {k} +(M+g_s\phi)^2}},
\end{equation}
and  is a suitable generalization
of the variable $\alpha$. 
The nucleon mass is taken to be $M+g_s\phi$, because   it is this  mass that
appears in the nucleon field equations.
The mass of the $A-1$ body system is dominated by the mass of the (A-1)
 nucleons mass 
(the binding energy per particle is 16 MeV $(\equiv \epsilon_B)$
compared with 940 MeV), so that we may
write
\begin{eqnarray}
\alpha_A&=&{\sqrt{\bbox{k}\cdot\bbox{k} +(M+g_s\phi)^2} +k^3\over
M_A}\left(1+ \epsilon_B/M_A+k_F^2/ 2(M+g_s\phi)M_A\right) \nonumber\\
&=& {\sqrt{\bbox{k}\cdot\bbox{k} +(M+g_s\phi)^2} +k^3\over
M_A},
\label{approx}
\end{eqnarray}
in which the last line 
results from the limit $A\to\infty$.
The key feature is that 
the variable $\alpha_A$ is defined as a momentum fraction, so that 
\begin{equation}\alpha_A M_A=k^+.\label{begin}
\end{equation}
Comparing Eq.~(\ref{approx}) and Eq.~(\ref{begin})
leads to Eq.~(\ref{kplus}).

The computation  of the energy and plus 
momentum distribution proceeds from taking the appropriate expectation
values of the energy 
momentum tensor $T^{\mu\nu}$ discussed in Sect.~2 and
\begin{equation}
P^\mu={1\over 2}\int d^2 x_\perp dx^- \langle T^{+\mu}\rangle.\label{pmu}
\end{equation}
We are concerned with the light front energy $P^-$ and momentum $P^+$.
The relevant components of 
$T^{\mu\nu}$ are presented in Eqs.~(\ref{tpm1}) and (\ref{tpp}).
Within the mean field approximation(MFA), the derivatives of the meson fields 
are  zero so that  one finds
\begin{eqnarray}
T^{+-}_{MFA}= m_s^2\phi^2 +2\psi_+^\dagger (i\partial^--g_v\bar V^-)\psi_+
\nonumber\\
T^{++}_{MFA}=m_v^2 V_0^2+ 2\psi^\dagger_+i\partial^+\psi_+.
\end{eqnarray}
 Taking the nuclear matter expectation
value of $T^{+-}$ and $T^{++}$ and performing the spatial integral of 
Eq. (\ref{pmu}) leads to the  result 
\begin{eqnarray}
{P^{-}\over \Omega}&=& m_s^2
\phi^2 +{4\over (2\pi)^3}\int_F d^2k_\perp dk^+ {\bbox{k}_\perp^2+ 
(M+g_s\phi)^2\over k^+}\label{pminus}\\
{P^{+}\over \Omega}&=& m_v^2 
V_0^2 +{4\over (2\pi)^3}\int_F d^2k_\perp dk^+ k^+.\label{pplus}
\end {eqnarray}
The subscript F denotes that $\mid\vec k\mid<k_F$   with $k^3$ defined
by the relation (\ref{kplus}).

Equations (\ref{pminus}) and (\ref{pplus}) along with the expression
for $k^+$, (\ref{kplus}) allow an evaluation of $P^-$ and $P^+$.
This shall be done in two different ways. In the first method we evaluate 
the energy of the A-nucleon system 
$E_A={1\over 2}(P^++P^-)$\cite{gs}, which turns out to be  
the  same as in the usual equal-time treatment\cite{bsjdw}. This can be seen by
summing equations (\ref{pminus}) and (\ref{pplus}) to obtain
\begin{eqnarray}
{E_A\over \Omega}= {1\over 2} m_s^2\phi^2 +
{1\over 2}m_v^2 V_0^2 
+{4\over (2\pi)^3}{1\over2}\int_F d^2k_\perp dk^+ \left({\bbox{k}_\perp^2+ 
(M+g_s\phi)^2\over k^+}+k^+\right)\label{Ei}
\end {eqnarray}
Then replace the integration over $k^+$ by one over $k^3$, using
Eq.~(\ref{kplus}) so that 
\begin{equation}
dk^+\to {k^+\over 
\sqrt{(M+g_s\phi)^2+ \bbox{k}\cdot\bbox{k}}}dk^3={k^+\over E(k)}dk^3.
\end{equation}
where
\begin{equation}
E(k)\equiv \sqrt{\bbox{k}\cdot\bbox{k}+
(M+g_s\phi)^2}.
\end{equation}
Then Eq.~(\ref{Ei}) takes the form:
\begin{equation}
{E_A\over \Omega}= {1\over 2} m_s^2\phi^2 +
{1\over 2} m_v^2 V_0^2
+{4\over (2\pi)^3}\int_F d^3k\theta(k_F-k)\;E(k),
\label{E}.
\end{equation}
which is the expression familiar from the Walecka model,
 this confluence
of energies is a nice check on the present result because  a manifestly  
covariant solution
 of the present problem, with the usual energy,  has been obtained\cite{sf}.

We consider the system to be at a fixed large volume, $\Omega$, so that
$E_A/A$ depends on $\phi$ and $k_F$. The ground state energy is determined 
by minimizing $E_A/A$ with respect to those two parameters.
Setting ${\partial E_A\over \partial \phi}$\cite{afac}
 to zero reproduces the field
equation for $\phi$ (\ref{smf})  as is also the case
in the equal-time formalism. 
The next step is to minimize  the energy per particle 
$E_A/A=E_A/(\rho_B\Omega)$  at
fixed volume with respect to $k_F$. (One may also 
minimize the energy with respect to the volume\cite{gs}.)
Start this calculation by using
\begin{equation}
{\partial\over \partial k_F}\left({E_A\over \rho_B}\right) =0
\end{equation}
to obtain:
\begin{equation}
{\partial E\over \partial k_F}=3{E\over k_F}.
\end{equation}
Using Eq.~(\ref{E}) followed by Eqs.~(\ref{smf}) and (\ref{mfa}) 
leads to the result
\begin{equation}
{4\over (2\pi)^3}{4\pi\over 3}k_f^3 E_F={m_s^2\over 2}\phi^2-{m_v^2\over 2}
V_0^2+
{4\over (2\pi)^3}\int_F d^3k\theta(k_F-k)E(k),
\label{mid}
\end{equation}
where $E_F\equiv E(k_F)$.
This is a transcendental equation which determines $k_F$, so
that the calculation of $E_A$ is complete. 

It is useful to  note that the relation
$P^+=P^-$ (which must hold for a system in its rest  frame) also emerges as
a result of this minimization. To see this rewrite 
the left hand side of 
Eq.~(\ref{mid}) as 
\begin{equation}
{4\over (2\pi)^3}{4\pi\over 3}k_f^3 E_F=
{4\over (2\pi)^3}
\int d^3k\theta(k_F-k)\left( E(k)+{\bbox{k}\cdot\bbox{k}\over 3 E(k)}\right).
\end{equation}
Using this in Eq.~(\ref{mid}) leads to 
\begin{equation}
{m_s^2\over 2}\phi^2-{m_v^2\over 2}
V_0^2= {4\over (2\pi)^3}\int
d^3k\theta(k_F-k){\bbox{k}\cdot\bbox{k}\over 3 E(k)},
\end{equation}
which is just the relation that one obtains by setting $P^+=P^-$ with
the versions of 
Eqs.~(\ref{pminus}) and (\ref{pplus}) obtained by replacing the variable
 $k^+$ by $k^3$.

Another way to obtain the energy of the ground state is to minimize
the value of $P^-/A$ subject to the constraint that $P^-=P^+$, or to minimize
the quantity ${\cal E}$ with
\begin{equation}
{\cal E}\equiv {P^-\over A}-\lambda \left({P^-\over A}-{P^+\over A}\right),
\end{equation}
where
$\lambda$ is a Lagrange multiplier. Setting 
$\partial{\cal E}\over\partial \phi$
leads to \cite{afac} 
\begin{equation}
{\partial P^-\over \partial \phi}(1-\lambda)
+\lambda {\partial P^+\over \partial \phi}=0 \label{mini}
\end{equation}
But the field equation (\ref{smf}) for $\phi$ can be restated as 
\begin{equation}
{\partial P^-\over \partial \phi}=-{\partial P^+\over \partial \phi}.
\label{rew}
\end{equation}
Combining Eqs.(\ref{mini}) and (\ref{rew}) leads to the result that
\begin{equation} \lambda={1\over 2},
\end{equation}
so that the minimization of $\cal E$ with respect to $k_F$ is the same
as minimizing $E_A/A$ with respect to $k_F$. This ends the discussion
of how the expressions for $P^\pm$ are used to determine the energy 
of the system.

Solving the  field equations and minimizing the energy density 
determines the properties of nuclear matter, once the 
meson-nucleon coupling constants and masses are chosen.
One can now discuss the properties of the resulting
system. Of course the necessary calculations have 
been done long ago. In particular, 
the parameters 
\begin{equation}
g_v^2M^2/m_v^2=195.9 \label{cv}
\end{equation} and  
\begin{equation}g_s^2M^2/m_s^2=267.1 \label{cc}
\end{equation} 
have been chosen
\cite{chin} so as to give the binding energy per particle of nuclear matter
as 15.75 MeV with $k_F$=1.42 Fm$^{-1}$. In this case, solving the 
equation for $\phi$ gives 
 $M+g_s\phi=0.56\;M$.

\subsection{Nucleon and Meson  Plus Momentum and Deep Inelastic Scattering}

The light front formalism embodies the  use of $k^+$ as a canonical variable
which allows us to study the nucleonic and mesonic contributions to the 
nuclear plus momentum. 
The study of the plus momentum content is motivated by the 
desire to obtain a better understanding of 
lepton-nucleus  deep inelastic scattering.
The EMC effect\cite{EMCrefs} that the 
structure function of a  bound nucleon differs from that of a free one,
showed a 
principal effect that the plus  momentum carried by the  valence quarks  is 
less for a bound nucleon than for a free one. 
Many different interesting  interpretations
and related experiments\cite{EMCrevs} were stimulated by these
experiments. But a correct  interpretation
requires that the role of
conventional effects, such as nuclear binding, be assessed and
understood.  

Our formalism employs  plus component of the momentum so that it's use in
assessing the nucleon's (and therefore the valence quark's) plus-momentum 
is necessary. 
We therefore  examine the $++$  component of the 
energy momentum tensor 
Eq.~(\ref{pplus}) to determine how much momentum is carried by 
nucleons and how much by mesons. 
Rewrite Eq.~(\ref{pplus}) as a sum of mesonic  $m$ and nucleonic $N$
terms 
\begin{equation}
{P^+\over A}={P^+_m\over A}+{P^+_N\over A},
\end{equation}
with 
\begin{equation}
{P^+_m\over A}={m_v^2 V_0^2 \over \rho_B} \label{pm}
\end{equation}
and
\begin{equation}
{P^+_N\over A}={4\over\rho_B (2\pi)^3}\int_F d^2k_\perp dk^+ k^+.\label{nuc}
\end{equation}
The parameters of Eqs.~(\ref{cv}) and (\ref{cc}) leads to 
\begin{equation}
{P^+\over A}=M-15.75 MeV,
\end{equation}
and the use of Eq.~(\ref{cv}) in Eq.~(\ref{pm}) gives:
\begin{equation}
{P^+_m\over A}=329 MeV,
\end{equation}
while performing the integral involved in Eq.~(\ref{nuc}) leads to
\begin{equation}
{P^+_N\over A}=594 MeV.
\end{equation}
The result is that 
65\% of the nuclear plus momentum
must be carried by the  nucleons  and 
the remainder of 35\% is carried by the  mesons.

How do these numbers relate to experiments? To answer we
need to recall that the nuclear structure function $F_{2A}$ 
can be obtained from the light front distribution function
$f(y)$ (which gives the probability for a nucleon to have 
a plus momentum fraction $y$) and the nucleon structure function
$F_{2N}$ using  the relation\cite{sfrel}:
\begin{equation}
{F_{2A}(x)\over A}=\int dy f(y) F_{2N}(x/y), \label{deep}
\end{equation}
where $y$ is the $A$ times the fraction of the 
nuclear  plus-momentum carried by the nucleon, and
$x$ is the Bjorken variable computed using the
nuclear mass divided by $A$ ($\bar M$):  $x=Q^2/2\bar M \nu$.
This formula  is the expression of the convolution model in which 
one means to assess, via  $f(y)$,  only  the influence of nuclear binding.
Other effects  such as the nuclear modification of the nucleon
structure function (if $F_{2N}$ is obtained 
from deep inelastic scattering on the
free nucleon) and 
any influence of the final state interaction between
the debris of the struck  nucleon and the residual nucleus\cite{st} are
neglected. 

Our formalism enables us to calculate
the function $f(y)$ from the integrand of 
Eq.(\ref{pplus}). Since the integral gives the total plus-momentum
carried by nucleons, the integrand which multiplies the
factor $k^+$ can be interpreted as  the necessary probability
distribution. Thus:
\begin{equation}P^+_N/A=\int
dk^+\;k^+ f(k^+),
\end{equation}
\begin{equation}
f(k^+)=\int_F d^2k_\perp.\label{fdef}
\end{equation} 
The function $f(y)$ can be obtained by 
replacing $k^+$ by the dimensionless variable $y$ using
$y\equiv {k^+\over \bar{M}}$
with $\bar{M}\equiv M-15.75 $ MeV. Then using Eq.(\ref{fdef})
leads to the result
\begin{equation}
f(y)={3\over 4} {\bar{M}^3\over k_F^3}\theta(y^+-y)\theta(y-y^-)\left[
{k_F^2\over \bar{M}^2}-({E_F\over \bar{M}}-y)^2\right], \label{fy}
\end{equation}
where
$y^\pm\equiv {E_F\pm k_F\over \bar{M}}$ and
$E_F\equiv\sqrt{k_F^2+(M+g_s\phi)^2}$. Knowing
the following integrals is useful:
\begin{eqnarray}
\int dy f(y)&=&1 \\
\int dy\; y\;f(y)&=&0.65,
\end{eqnarray}
with the 0.65 representing the earlier 65\% result.
 
We may now assess the implications of the statement that nucleons carry
only 65\% of the nuclear plus-momentum. This number is to be compared
with the value obtained by
Frankfurt and Strikman\cite{fs2}.   
They used data for $F_{2A}$ and $F_{2N}$ along with Eq.~(\ref{deep})
to determine the average value of $y$ 
required by experiments with the result that
\begin{equation}
\int dy\; y\;f_{exp}(y) =0.95.
\end{equation}
This means that  
nucleons carrying   95\% of the nuclear plus momentum 
(a 5\% depletion effect)
is sufficient to explain the 10-15\% 
depletion effect observed for  the Fe nucleus. 
Our 35\% depletion seems to be rather large, but one must remember
that it results from nuclear matter.
A result that compares more closely with experiment could be obtained
in a version of the present model for which the value of $M+g_s\phi$ is closer
to $M$. However, determining specific features of the present model
is not the goal of the present work. Instead we  wish to demonstrate that
the light front formalism can be used to obtain a nuclear wave function
expressed in terms of the plus-momentum variable which is closely related
to experiment.

Indeed we can verify that the ability to obtain the nucleonic plus 
momentum,  feature that requires the use of a 
light front front formalism, instead of an equal time formalism. To do
this compare the 0.65 fraction with the result of a
relativistic calculation using the equal time (et)
formalism\cite{birse}.  In this calculation, which uses
Eq.~(\ref{lag})) and for which the scalar and vector fields are the
same as here, the plus momentum of a nucleon was chosen as the sum of
the Dirac eigenenergy and $k^3$:
\begin{equation}
k^+_{et}\equiv E_{Dirac}+k^3=\sqrt{(M+g_s\phi)^2+\bbox{k}^2}+g_vV^0 +
k^3. \label{ident}
\end{equation}
Using this leads to an average nucleon plus momentum fraction 
$\langle y \rangle_{et}= (E_F+g_v V^0)/\bar{M}$, which when evaluated with our
parameters  for $k_F,\phi$ and $\bar V^-$, leads to $\langle y
\rangle_{et}= 1.00$!  The big difference between our result and the
earlier equal time result -compare Eqs.~(\ref{kplus}) and
(\ref{ident})- arises from our use of the plus momentum as a
canonical momentum variable and the consequent use of $T^{+\mu}$ to
construct the light front momentum and energy density.
In particular, the  first line of Eq.(\ref{ident}) is only a reasonable guess.

We note also that  the baryon number distribution $f_B(y)$
 (number of baryons per $y$, normalized
to unity) can be
determined from the expectation value of $\psi^\dagger\psi$. The result is
\begin{equation}
f_B(y)={3\over 8} {\bar{M}^3\over k_F^3}\theta(y^+-y)\theta(y-y^-)
  \left[ (1+{E_F^2\over \bar{M}^2y^2})    (
 {k_F^2\over \bar{M}^2}-({E_F\over \bar{M}}-y)^2)
 -{1\over 2y^2}({k_F^4\over \bar {M}^4}-({E_F\over \bar{M}}-y)^4)
   \right].\label{fby}
\end{equation} 
Some phenomenological models  treat the two distributions 
$f(y)$ and $f_B(y)$ as identical. The distributions have 
the same normalization,
 but they are different as  shown by Eqs.~(\ref{fy}) and (\ref{fby}).

Consider that the average value of 
$y$ equal to 0.65 represents  
a very strong binding effect on lepton-nucleus deep inelastic scattering.
One might  think that the mesons, 
which cause this binding, would also have huge effects on
deep inelastic scattering.
It is therefore  certainly necessary to 
determine the  momentum distributions of the  mesons.
The mesons contribute 0.35 of the total 
nuclear plus momentum, but we need to 
know how this  is distributed over different individual values.
The paramount feature is 
that $\phi$ and $V^\mu$ are the same constants for any  and all 
values  of the spatial coordinates $x^-,\bbox{x}_\perp$ (and also $x^+$).
This means that
the related momentum distribution can only be proportional to a delta function
setting both the plus and $\perp$ components of the momentum to zero.
This result is attributed to the mean field approximation for infinite
nuclear matter, in which the meson
fields are treated as classical quantitates. Thus the finite plus momentum 
can be thought of as coming from an infinite number of quanta, each carrying
an infinitesimal amount of plus  momentum. A plus momentum of 0 can only be
accessed experimentally at $x_{Bj}=0$, which requires an infinite amount 
of energy. Thus, in the mean field approximation,
the scalar and vector mesons can not contribute to deep 
inelastic scattering. The usual term  for   a
field  that is constant over space
is a zero mode, and  the present Lagrangian provides a simple example.
For finite nuclei, in the mean field approximation,
the mesons would carry a very small momentum of scale given
by the inverse of the nuclear radius, under
the mean field approximation. If fluctuations were to be included, the 
relevant momentum scale would be  of the order of 
the inverse of the average distance between
nucleons (about 2 Fm).

We can understand   the significance of the presence of
components of a wave function
that carry plus momentum but do not participate
deep inelastic processes by reviewing a bit of history.
The nuclear binding effect is that 
the plus momentum of a bound nucleon is reduced by the binding
energy, and so is that of its confined quarks.  Conservation of
momentum implies that if nucleons lose momentum, other constituents
such as nuclear pions\cite{ET}, must gain momentum. This partitioning
of the total plus momentum amongst the various constituents is called
the momentum sum rule.
Pions are quark anti-quark pairs so that a specific enhancement of
the nuclear antiquark momentum distribution, mandated by momentum
conservation, is a testable \cite{dyth} consequence of this idea.  A
nuclear Drell Yan experiment \cite{dyexp}, in which a quark from a
beam proton annihilates with a nuclear antiquark to form a
$\mu^+\mu^-$ pair, was performed. No influence of nuclear pion
enhancement was seen, leading Bertsch et al.\cite{missing} to question
the idea that the pion is a dominant carrier of the nuclear force.
In the present situation, we have a huge depletion effect of 35\%, but
with no consequence for either the nuclear deep inelastic scattering or
Drell-Yan experiments.

We hasten to add that the Lagrangian of Eq.~(\ref{lag}) and its
evaluation in mean field approximation for nuclear matter have been
used to provide a simple but semi-realistic example.  It would be
premature to compare the present results with data before obtaining
light front dynamics for a model in which the correlational
corrections to the mean field approximation are included, and which
treats finite nuclei.  Thus the specific numerical results of the
present work are far less relevant than the central feature that the
mesons responsible for nuclear binding need not be accessible in deep
inelastic scattering.

However, the present model may be regarded as being one of 
a class of models in which the mean field plays an important role\cite{qmc}. 
For such models  
nuclei would have constituents that contribute
to  the momentum sum rule but do not
contribute to deep inelastic scattering.
In particular, a model can have a large binding effect, nucleons can carry a
significantly less fraction of $P^+$ than unity,  and it is not necessary to
include the influence of mesons that could be  ruled out 
in a Drell-Yan experiment.

\section{Summary and Discussion}
The present paper shows how the light front quantization of a chiral
Lagrangian can be accomplished.  The resulting formalism can be
applied to many problems of interest to nuclear physics. 
In particular, 
pion-nucleon,  nucleon-nucleon
scattering, and infinite nuclear matter (in the mean field approximation)
are presented here.   Soft pion theorems for pion-nucleon scattering are
reproduced.  The treatment of nucleon-nucleon scattering is shown to
be manifestly covariant in the one boson exchange approximation. The
implications of chiral symmetry for the two-nucleon intermediate state
contribution to the two pion exchange potential are discussed.  The
present results mainly constitute a feasibility study, in that the
emphasis here is on checking the formalism by reproducing known
results.  But this light front treatment does allow the effects of
chiral symmetry to be incorporated within a relativistic formalism,
and therefore should have a broad applicability in the future. One
remaining technical problem is to provide a light front quantization
of a Lagrangian for spin 3/2 particles.

The special feature of the light front formalism is its
use of the plus momentum as one of the canonical variables. This enables
a close contact with the experimental variables used to analyse 
deep inelastic scattering and any experiment in which there is one large
momentum. This feature is exploited here in the derivation 
(within the mean field approximation) of the 
nucleonic and mesonic distribution functions for infinite nuclear
matter. The mesons are shown to
carry a significant fraction of the nuclear plus momentum, but only 
a zero plus-momentum (a zero mode), and therefore do not
participate in nuclear deep inelastic scattering or Drell-Yan experiments. 

The ultimate validity of the above (perhaps startling) statement
depends on whether or not the dominance of mesonic zero modes survives
calculations performed for finite nuclei and calculations which
include the effects of nucleon-nucleon correlations. There seems to be
no technical barrier precluding such calculations.

\acknowledgements
This work is partially supported by the USDOE. I thank the SLAC theory group
and the national INT for their hospitality.
I  thank P. Blunden, S.J. Brodsky, L. Frankfurt, 
S. Glazek, R. Machleidt, 
R.J. Perry, C.M. Shakin,   M. Strikman and U. van Kolck
for useful discussions.

\appendix
\section{Notation, conventions, and useful relations}
\noindent This is patterned after the review of Harindrinath\cite{hari}
 The  light-front variables are defined by
\begin{equation} x^{+}= x^{0}+x^{3} \; , \; \;\; x^{-}= x^{0}-x^{3},
\label{def}
 \end{equation}
so  the four-vector $x^\mu$ is denoted
\begin{equation} x^{\mu} = (x^{+},x^{-},\bbox{x}^{\perp}) . \end{equation}
With this notation the 
scalar product is denoted by 
 \begin{equation} x\cdot y = {1 \over 2} x^{+}y^{-}+{1 \over 2}x^{-}y^{+}-
\bbox{x}^{\perp}\cdot \bbox{y}^{\perp}  . \end{equation}
The metric tensor $g^{\mu \nu}$ with $\mu=(+,-,1,2)$ is obtained from the 
usual one by using (\ref{def}) (i.e. $g^{0\mu}=g^{0\mu}+g^{3\mu}$). Then 
$g^{+-}=g^{-+}=2,g^{ij}=-1$, with the other elements vanishing.
The term $g_{\mu\nu}$ is obtained from the condition  that 
$g^{\alpha\beta}g_{\beta\gamma}=\delta_{\alpha\gamma}$. Its elements are the
same as those of $g^{\mu\nu}$ except for $g_{-+}=g_{+-}=1/2$.
Thus
  \begin{eqnarray}
 x_{-}= {1 \over 2} x^{+}  , \; \; x_{+} = {1 \over 2} x^{-}, \end{eqnarray}
\noindent and the  partial derivatives are similarly given by 
\begin{eqnarray}
\partial^{+}= 2 \partial_{-}= 2 {\partial \over \partial x^{-}} \;\quad
\partial^{-}= 2 \partial_{+}= 2 {\partial \over \partial x^{+}}  . 
\end{eqnarray}

 The step function is defined as  $ \theta (x) =1$ for $x>0$,
and $ \theta (x) =0$ for $x\le0$.
The antisymmetric step function is given by 
\begin{eqnarray}
\epsilon(x) = \theta(x) - \theta(-x)  , \end{eqnarray}
 with
\begin{eqnarray} {\partial \epsilon \over \partial x} \; \; =
2\;\delta(x).\end{eqnarray}
In this notation, $
 \mid x \mid \; \; = \; \; x \; \epsilon(x)  $. The above few definitions
allow us to express the inverse operators appearing in the text in terms of
integrals: 
\begin{eqnarray} {1 \over \partial^{+}} f(x^{-})= {1 \over 4}\, \int \, dy^{-}
\epsilon(x^{-}-y^{-}) \, f(y^{-})  ,\label{inv} \end{eqnarray}
\begin{eqnarray}
  ({1 \over \partial^{+}})^{2} f(x^{-}) =  { 1 \over 8} \, \int \,
dy^{-} \mid x^{-} - y^{-} \mid \, f(y^{-})  . \end{eqnarray}
\noindent  

The Bjorken and Drell\cite{Bj}  convention for gamma matrices is used
and 
\begin{eqnarray} 
\gamma^{\pm} \; \; \equiv \gamma^{0} \pm \gamma^{3}  . \end{eqnarray}
The relations 
\begin{equation}\gamma^\pm\gamma^\pm=0,
\quad\gamma^+\gamma^-\gamma^+=4 \gamma^+,\quad 
\gamma^-\gamma^+\gamma^-=4 \gamma^-
\end{equation} can be used to simplify various computations.

The hermitian projection operators $\Lambda_\pm$ are given by 
\begin{eqnarray}
  \Lambda_{\pm}  \; =  \; {1 \over 4} \gamma^{\mp} \gamma^{\pm}
 = {1 \over 2} \gamma^{0} \gamma^{\pm} \; = \; {1 \over 2}
(I \pm \alpha^{3})  , \end{eqnarray}
and obey the following relations
\begin{eqnarray} (\Lambda_{\pm})^2  \; \; = \; \; \Lambda_{\pm},\quad \quad
\gamma^{\perp} \; \Lambda_{\pm}, \; \; 
= \; \; \Lambda_{\pm} \gamma^{\perp}  , \end{eqnarray}
\begin{eqnarray}
  \gamma^{0} \; \Lambda{\pm} \; \; = \; \; \Lambda_{\mp} \gamma^{0}
\quad\quad
 	   \alpha^{\perp} \; \Lambda_{\pm} \; \; 
= \; \; \Lambda_{\mp} \alpha^{\perp} , \end{eqnarray}
\begin{eqnarray}  \gamma^{5} \; \Lambda_{\pm} \; \; 
= \; \; \Lambda_{\pm} \gamma^{5} \quad
 \gamma^\mp = 2 \Lambda_\pm \gamma^0 = \gamma^\mp \Lambda_\mp ,\end{eqnarray}
\begin{eqnarray}
 \gamma^i \Lambda_\mp = {1 \over 2} \gamma^i \pm i {1 \over 2} \epsilon^{ij}
\gamma^j \gamma^5 , \end{eqnarray}
and
\begin{eqnarray}
 \alpha^j \gamma^i \Lambda_+ = {i \over 2} \epsilon^{ij} \gamma^+ \gamma^5
. \end{eqnarray}   
\noindent The  Dirac spinors are given by 
\begin{eqnarray} u_{\lambda}(k) = \sqrt{2 \over k^{+}} \big [ M \; \Lambda_{-}
\; \; + \; \; (k^{+}\; + \; \bbox{\alpha}^{\perp}\cdot\bbox{k}^{\perp}) 
\; \Lambda_{+}\big ] \; \chi_{\lambda}  ,\end{eqnarray}
where
\begin{eqnarray} \chi^\dagger _{\uparrow} = (1,0,0,0), \quad
  \chi^\dagger_{\downarrow} = (0,1,0,0). \end{eqnarray}
The anti-particle spinors are given by 
$  v_{\lambda}(k) = C \; ({\bar u_{\lambda}(k)})^{T} \; $
where $C = i \gamma^{2} \gamma^{0}$ is the charge conjugation operator, so that
\begin{eqnarray}
  v_{\lambda}(k) = \sqrt{2 \over  k^{+}} \big [ M \; \Lambda_{-}
\; \; + \; \; (k^{+}\; + \; \bbox{\alpha}^{\perp}\cdot\bbox{k}^{\perp}) \; 
\Lambda_{+}\big ] \; \eta_{\lambda}  \end{eqnarray} with
\begin{eqnarray}  \eta^\dagger_{\uparrow} = (0,0,0,1), \quad
  \eta^\dagger_{\downarrow} = (0,0,-1,0). \end{eqnarray}
Other useful relations are 
\begin{eqnarray}
\bar u(\bbox{k},\lambda)u(\bbox{k},\lambda')=2M \delta_{\lambda,\lambda'},
\quad
\bar v(\bbox{k},\lambda)v\bbox{k},\lambda')=-2M \delta_{\lambda,\lambda'},
\end{eqnarray}
\begin{eqnarray}
\bar u(\bbox{k},\lambda)\gamma^\mu
u(\bbox{k},\lambda')=2p^\mu\delta_{\lambda,\lambda'}, \quad
\bar v(\bbox{k},\lambda)\gamma^\mu v\bbox{k},\lambda')=-2p^\mu
 \delta_{\lambda,\lambda'},
\end{eqnarray}
and
\begin{eqnarray}
\sum_\lambda u(\bbox{k},\lambda)\bar u(\bbox{k},\lambda) =\gamma\cdot k+m,
\quad
\sum_\lambda v(\bbox{k},\lambda)\bar v(\bbox{k},\lambda) =\gamma\cdot k-m.
\end{eqnarray}
Note that in the above three  equations $k^\mu$ is an on-shell four vector
with $k^-={\bbox{k}_\perp^2+M^2\over k^+}$ and $\bbox{k}=(k^+,\bbox{k}_\perp)$.

\newpage
{\bf Figure Captions} 
\begin{figure}[t]
\noindent
\epsfysize=4.5in
\hspace{1.0in}
\epsffile{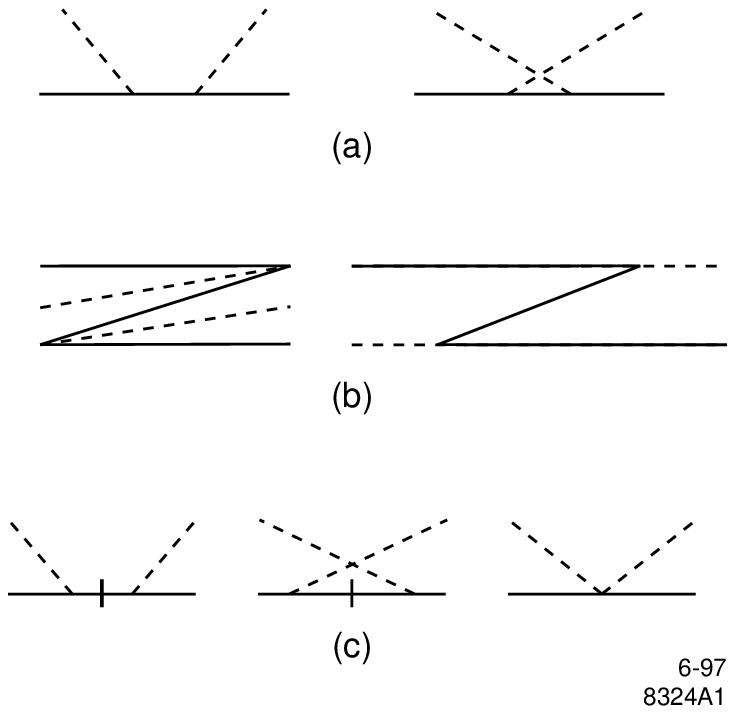}
\begin{center}
Fig.~1 $x^+$-ordered graphs for low energy pion-nucleon scattering.
(a) Second-order effects of the $\bar u \gamma_5 u$ term $v_1$. 
(b) Second-order effects of the $\bar u \gamma_5 v$ and  $\bar v \gamma_5 u$
 terms of  $v_1$. (c) Effects of the instantaneous fermion propagation terms
of $v_2$, and of the $\pi^2$ term of $v_1$. The terms $v_i$ are defined
in Eqs,~(2.46)-(2.48).
\end{center}
\end{figure}
\begin{figure}[t]
\noindent
\epsfysize=4.5in
\hspace{1.0in}
\epsffile{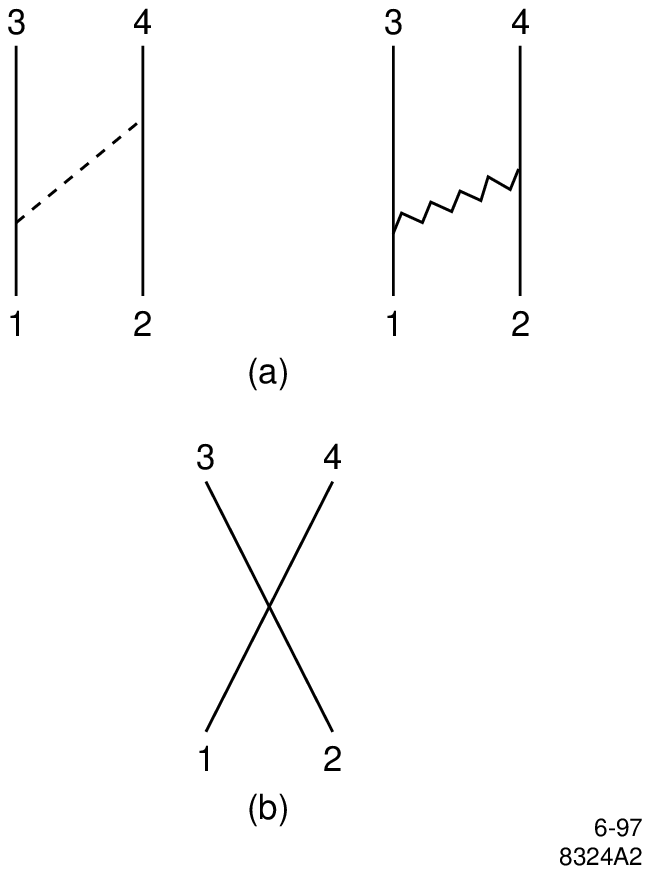}
\begin{center}
Fig.~2. $x^+$-ordered graphs for one boson exchange contributions 
to nucleon-nucleon scattering. The numbers 1-4 represent the momentum, spin and
charge states 
of the nucleons. Here $k^+_1>k^+_3$. (a) meson propagation terms 
(b) instantaneous  vector meson exchange of $v_3$, Eq.~(2.48)
\end{center}
\end{figure}
\newpage 
\begin{figure}[t] 
\noindent
\epsfysize=4.5in
\hspace{1.0in}
\epsffile{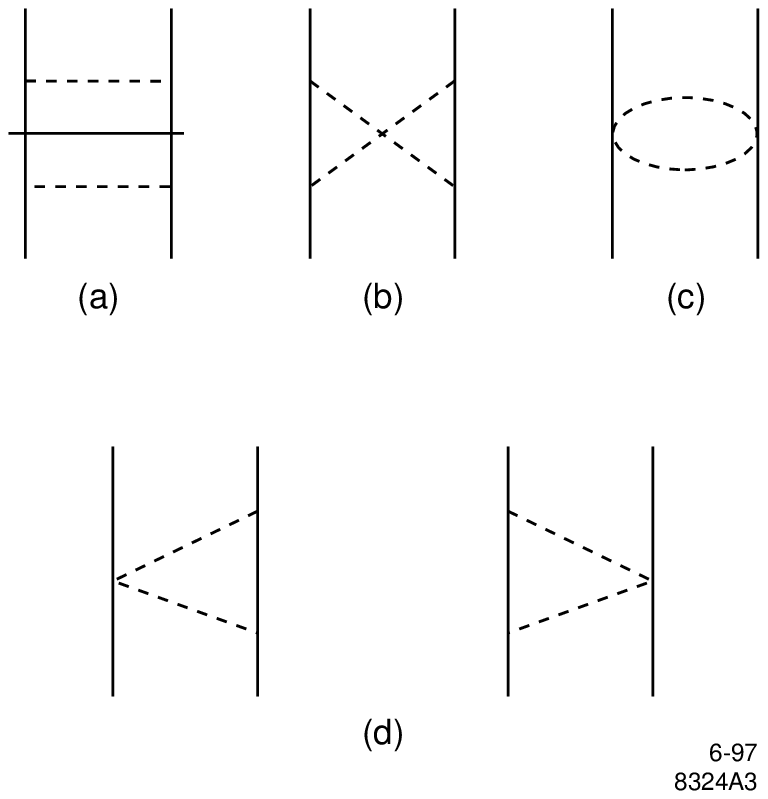}
\begin{center}
Fig.~3
Feynman graphs for the two-pion exchange potential
(a) uncrossed box diagram- the horizontal line represents the subtraction of
the contribution arising from the iterated one pion exchange potential.
(b) Crossed box diagram (c) Second order effect of the $\pi^2$ term of
$v_1$, Eq.~(2.46) (d) Terms with one $\pi^2$ term.
\end{center}
\end{figure}


\begin{thebibliography}{99}
\bibitem{notation} Our notation is that a four vector $A^\mu$ is defined by
the plus, minus and perpendicular components as
$(A^0+A^3,A^0-A^3,\bbox{A}_\perp)$ and $A\cdot B={1\over 2} A^+B^-+{1\over
2}A^-B^+-\bbox{A_\perp\cdot B_\perp}$. See the Appendix for more on this.
\bibitem{scalar} The use of scalar mesons is meant as a simple  representation
of part of the two pion
exchange potential which causes much of the  
the medium range attraction between nucleons, as well as the effects of 
a possible fundamental scalar meson. 
\bibitem{bira} The most general non-linear realization does not specify
the ratio of the two pion-nucleon couplings.
\bibitem{exp} The symbol for the nucleon field operator and the mode functions
of that field  is taken to be the same -$\psi$ to reduce the amount 
of notation.
\bibitem{mini} Minimal substitution would require that
the plus component of the vector field would
enter into the equation (\ref{kplus}) for $k^+$, but we are using
a formalism for which that component vanishes.
\bibitem{afac} Minimizing $E_A$ with respect to $\phi$ (for fixed $k_F$)
is the same as
minimizing $E_A/A$ with respect to $\phi$  (for fixed $k_F$) because 
the volume $\Omega$ is held fixed.
\bibitem{sfrel} Eq.~(\ref{deep}) is essentially Eq. (5.2) of 
Frankfurt and Strikman, \cite{fs2}, with a correspondence
between our $A\;f(y)$ 
and their $\int d^2p_\perp \rho^N_A(\alpha,p_\perp)/\alpha$.
\end{thebibliography}
\end{document}